# EIGENRAYS IN 3D HETEROGENEOUS ANISOTROPIC MEDIA:

# PART V – DYNAMICS, VARIATIONAL FORMULATION

*Igor Ravve (corresponding author) and Zvi Koren, Emerson*

igor.ravve@emerson.com , zvi.koren@emerson.com

## ABSTRACT

Parts V, VI and VII of this study are dedicated to the computation of paraxial rays and dynamic characteristics along the stationary rays obtained numerically in Part III. In this part (Part V), we formulate the linear, second-order, Jacobi dynamic ray tracing equation. In Part VI, we compare the Lagrangian and Hamiltonian approaches to the dynamic ray tracing in heterogeneous isotropic and anisotropic media; we demonstrate that the two approaches are compatible and derive the relationships between the Lagrangian's and Hamiltonian's Hessian matrices. In Part VII, we apply a similar finite-element solver, as used for the kinematic ray tracing, to compute the dynamic characteristics between the source and any point along the ray. The dynamic characteristics in our study include the relative geometric spreading and the phase correction factor due to caustics (i.e., the amplitude and the phase of the Green's function of waves propagating in 3D heterogeneous general anisotropic elastic media).

The basic solution of the Jacobi equation is a shift vector of a paraxial ray in the plane normal to the ray direction at each point along the central ray. A general paraxial ray is defined by a linear combination of up to four basic vector solutions, each corresponds to specific initial conditions related to the ray coordinates at the source. We define the four basic solutions with two pairs of initial condition sets: point-source and plane-wave. For the proposed point-source ray coordinates and initial conditions, we derive the ray Jacobian and relate it to the relative geometric spreading for general anisotropy.

Finally, we introduce a new dynamic parameter, referred as the “normalized” geometrical spreading, which can be considered a measure of complexity of the propagated wave/ray phenomena, and hence, we propose using a criterion based on this parameter as a qualifying factor associated with the given ray solution.

Keywords: General anisotropy, Paraxial rays, Geometric spreading, Caustics, KMAH index.

## INTRODUCTION

Following the theory and implementation of the kinematic ray tracing, in smooth heterogeneous anisotropic elastic media, with the Eigenray method, described in Parts I - III of this study, Parts V - VII are devoted to the formulation and implementation of the corresponding dynamic ray tracing problem, respectively. The dynamic properties are the high-frequency characteristics of body waves propagating along precomputed stationary rays. The same finite-element scheme (including discretization and interpolation type), applied in Part III for the kinematic ray tracing (KRT), is used in Part VII to solve the corresponding dynamic ray tracing (DRT) vector-form ordinary differential equation (ODE) (referred as the Jacobi DRT equation). Solving the Jacobi DRT equation makes it possible to accurately compute the relative geometric spreading with special attention to the identification and classification of caustics.

Recall that in Part IV of this study we describe an efficient method for computing the global dynamic characteristic – the relative geometric spreading between the two endpoints of the stationary path, using the traveltime Hessian. Although that method doesn’t require explicitly solving the DRT equation, it doesn't deliver the relative geometric spreading at any point along the ray path, possible caustic locations, and their types; this requires performing the DRT explicitly.

For simplicity, from now on and throughout the paper, we use the term "geometric spreading" for the so-called "relative geometric spreading" defined, for example, by Červený (2000).

The proposed Eigenray method is a finite-element approach, primarily designed to solve the two-point (nonlinear) KRT problem and then to perform the (linear) DRT to obtain dynamic characteristics along the stationary ray. It is in particular valuable in complex geological areas, with considerable spatial variabilities of the anisotropic elastic properties, where conventional ray tracing becomes challenging. The KRT provides a stationary ray between two fixed endpoints, given an initial guess trajectory, where the kinematic characteristics are the nodal spatial coordinates, slowness and ray velocity vectors, and traveltime. Obtaining the dynamic characteristics involves approximations of paraxial rays (or a ray tube) for different initial/boundary conditions and, in particular in this study, computation of the point-source ray Jacobian (the varying, signed normal cross-section area of the ray tube) which makes it possible to analyze caustic locations and their types, and to compute the geometric spreading. The computation of these dynamic properties in complex geological areas (involving considerable heterogeneity and anisotropy effects), in particular across caustic locations, is challenging and can be unstable. The computation of the kinematic and dynamic properties requires the first and second spatial and directional derivatives of the traveltime, which can be formulated analytically in terms of the corresponding derivatives of the local ray (group) velocity vector (Ravve and Koren, 2019). Note that the actual computation of these derivatives requires stable numerical (first and second) derivatives of the stiffness (elastic) tensor components with respect to (wrt) the location coordinates.

The propagation of high-frequency energy within subsurface geological models may involve caustic locations along the rays. Caustics are associated with the zeros of the ray Jacobian along the ray path, and a simple criterion allows distinguishing between their two possible types (orders): line and point. In the case of a line caustic, we also establish the line direction. The (reciprocal) geometric spreading is directly related to the (non-reciprocal) ray Jacobian. Note also that the ray Jacobian depends on the choice of the ray coordinates (RC) at the source, while the geometric spreading is an objective physical characteristic independent of this choice. For isotropic media, the relationship between the ray Jacobian and the geometric spreading is very simple, which is not the case for anisotropic media. Using our proposed ray coordinates (RC), we provide a relatively simple relationship between the ray Jacobian and the (reciprocal, relative) geometric spreading in general anisotropic media. Generally, the ray tube used for computing the ray Jacobian is defined with two point-source DRT solutions, governed by a linear, vector-form, ordinary differential equation (ODE). The DRT equation provides the changes of the spatial coordinates, ray directions and slowness vectors of paraxial rays along the central ray wrt the RC (normally defined at the source location). The DRT is normally computed either in Cartesian coordinates, or in ray-centered coordinates (RCC), or in wavefront orthonormal coordinates (WOC) (e.g., Červený, 2000).

We introduce a suitable dynamic characteristic, which we call the normalized relative geometric spreading, $L_{GS}/\sigma$. This parameter is identically 1 along the ray paths in isotropic media of constant velocity gradient, and becomes different otherwise. At the source point, both the numerator and the denominator of this fraction vanish; however, their ratio has a finite limit. For arbitrary isotropic media, the limit is 1; for anisotropic media it accepts a different value, and we derive the formula to compute this limit. We find this new dynamic characteristic an attractive parameter to qualify the complexity of the wave phenomena along the ray. In some modeling, imaging or inversion (tomography) applications, this parameter can be used either as a threshold, for accepting/rejecting the obtained stationary path solutions, or as a weighting factor, for example, in the amplitude summation process in ray-based Kirchhoff-type migrations. Next, we derive the slope of this function: the arclength derivative of the normalized relative geometric spreading. The analytical formulae for these two values, $\left.\frac{L_{GS}}{\sigma}\right|_S$ and $\left.\frac{d}{ds}\frac{L_{GS}}{\sigma}\right|_S$, are useful for testing of the numerical results, delivered by the finite-element solutions.

In this part of our study we follow the main ideas and concepts, published more than one hundred years ago by Bliss (1916), on the Jacobi condition for variation calculus of parametric functionals. We adjust the general theory presented in this paper in order to derive the linear, second-order DRT equation, and name it the Jacobi DRT equation. The formulation is based on the analysis of the second traveltime variation of a general paraxial ray in smooth heterogeneous general anisotropic elastic media.

In Part VI, we compare the proposed Lagrangian DRT approach with a more common Hamiltonian approach and demonstrate that the two methods are fully compatible. We also derive the relationships between the Hessians of the Lagrangian, $L(\mathbf{x},\mathbf{r})$, wrt the ray location and direction vectors and the corresponding Hessians of its matching Hamiltonian, $H(\mathbf{x},\mathbf{p})$, wrt the ray location and slowness vectors.

In Part VII, we use the same finite-element approach applied in Part III to the kinematic Euler-Lagrange KRT equation, to solve the Jacobi DRT equation. We apply the weak finite-element formulation with the Galerkin method, to this second-order DRT ODE, thus effectively reducing it locally to a linear, first-order, local weighted residual equation set. In the finite-element implementation for the kinematic ray tracing (Part III), the resulting algebraic equation set is nonlinear, while in the implementation for the dynamic ray tracing (Part VII), it is linear. Due to the unique formulation of the proposed finite-element approach (discretization of the ray path and interpolation between the nodes for both the kinematic and dynamic analyses), the resolving matrix of the dynamic finite-element linear equation set (the "stiffness" matrix) becomes identical to the global (all-node) traveltime Hessian matrix computed for the stationary path in the kinematic ray tracing. This is a great advantage which makes the combined kinematic and dynamic computations naturally consistent and very efficient (since the "stiffness" matrix: the traveltime Hessian, is already available). In other words, the global traveltime Hessian computed for obtaining the stationary path inherently provides the "stiffness" matrix of the finite-element solver for computing the dynamic characteristics.

Structure of the paper

In the Introduction of Part IV we have reviewed the published studies describing the computation of the geometric spreading between given source and receiver locations, in particular, methods based on the source-receiver traveltime Hessian matrix, which do not require explicitly performing the DRT. In this part (Part V) we review the published works for computing dynamic properties along the ray path using the dynamic ray tracing. These methods are based on the computation of the ray Jacobian (i.e., the varying, signed normal cross-section area of the ray tube, representing also the determinant of the transformation matrix between the Cartesian coordinates and the RC of a paraxial ray), which, in turn, makes it possible to compute the geometric spreading and to identify caustic locations and their types (orders).

We then review the properties (conditions) of a stationary solution for a multi-dimensional parametric functional, studied by Bliss (1916), and we adjust the general theory in the cited paper to our case: The analysis of dynamic properties of rays in smooth heterogeneous general anisotropic elastic media. We then discuss the necessary and sufficient conditions for distinguishing between the two possible stationary ray solutions: A minimum traveltime, which requires a positive second traveltime variation, and a saddle point (due to caustics), where the second traveltime variation is not necessarily positive. Note that alternatively, one can conclude about the type of the stationary path by directly analyzing the traveltime Hessian of this path obtained in the kinematic stage: Positive definite traveltime Hessian means a minimum traveltime, while indefinite Hessian (with eigenvalues of both signs) is an evidence of a saddle point stationary solution.

Using variational calculus, we derive the Jacobi DRT equation. The solution to the Jacobi DRT equation is a vector, defined at each point along the central ray, representing a shift of a paraxial ray in the plane normal to the direction of the central ray at that point. In the most general case, this normal shift vector is obtained as a linear combination of (up to four) basic vector solutions (e.g., corresponding to two IC pairs: two point-source and two plane-wave initial condition sets) of the Jacobi DRT equation. We explicitly define the corresponding two point-source and the two plane-wave ray coordinates (RC). For the two point-source basic vector solutions, we formulate the corresponding ray Jacobian; its analysis provides the identification of caustic locations and characterization of their types. We then provide an original relation between the ray Jacobian and the relative geometric spreading for general anisotropy. A general paraxial ray consists of all four basic solutions is then used to relate the vector shifts imposed at the endpoints of the ray (source and receiver) to the four ray coordinates and to provide the corresponding paraxial traveltime. Finally, we compare the proposed Lagrangian and the corresponding Hamiltonian approaches to the dynamic ray tracing and demonstrate that they are fully compatible/equivalent for general anisotropy.

Appendices

In order to make the paper more readable, the body of the paper contains the main concepts of the proposed Lagrangian-based Jacobi DRT approach, with the principal governing equations and numerical examples, with minimum mathematical derivations. The detailed derivations have been moved to the appendices.

In Appendix A, we derive the arclength-dependent coefficients of the proposed linear Jacobi DRT equation, which are the first and second derivatives of the traveltime integrand (the Lagrangian) $L$ wrt the location and direction of the ray trajectory nodes.

In Appendix B, we study the necessary and sufficient conditions for a minimum traveltime stationary ray path. This is primarily important for the identification of caustics.

In Appendix C, we consider the second traveltime variation and derive the linear Jacobi DRT equation, that includes the coefficients obtained in Appendix A.

In Appendix D, we construct the initial conditions for both source-point and plane-wave paraxial rays.

In Appendix E, we relate small source/receiver vector shifts $\Delta\mathbf{x}_S$ and $\Delta\mathbf{x}_R$ to the required four ray coordinates (RC), $\gamma_i$ , $i=1,\ldots 4$, consisting of two point source RC $\gamma_i$ , $i=1,2$ and two plane wave RC $\gamma_i$ , $i=3,4$.

In Appendix F, we derive the ray Jacobian and the corresponding caustic criteria. We then provide the relationship between the (proposed) ray Jacobian and the (reciprocal) geometric spreading in general anisotropic media. The latter is commonly used as the amplitude factor of the Green's function.

In Appendix G, we establish the limit for the normalized relative geometric spreading $L_{GS}/\sigma$ of a point-source paraxial ray at the start point, and the limit for its arclength derivative (the slope of the normalized spreading).

## DRT-BASED RAY JACOBIAN AND GEOMETRIC SPREADING: A REVIEW

The use of the DRT-based ray Jacobian provides the computation of the corresponding geometrical spreading and the analysis of caustics at points along the stationary ray path. Neighboring curved rays associated with the same seismic event can come together, or focus, to form an envelope or a caustic surface. In these locations along the ray, the cross-section area of the ray tube shrinks to zero, leading to anomalously large amplitudes and specific phase shifts of Green's functions, which are fundamental components of many physical problems. These caustic-based phase shifts are normally associated with the KMAH index, named after the pioneering studies by Keller (1958), Maslov (1965), Arnold (1967) and Hörmander (1971). The caustics phenomena have been since widely investigated in many different wave/ray-based physical problems (e.g., Bott, 1982; Nye, 1985; White et al., 1988; Holm, 2012; Gutenberg and Mieghem, 2016; Červený, 2000, 2013).

Hubral et al. (1995a, 1995b) developed a new factorization method to establish geometric spreading and the number of caustics (KMAH index) along the ray path. With this approach, integrations along the ray segments can be performed independently; the global geometric spreading and the KMAH index then include contributions of the ray segments and Fresnel zones at the segment connections.

For 3D inhomogeneous general anisotropic elastic media, Gajewski and Pšenčík (1987) suggested a finite-difference approach to establish the components of the transform matrix $\mathbf{Q}$ where the derivatives of the Cartesian coordinates wrt the ray coordinates are computed for a fixed traveltime. Garmany (2000) generalized the KMAH sign rule for caustics on convex and concave quasi-shear slowness surfaces. In the latter case, the curvatures of the slowness surface should be computed in order to draw conclusions about its convexity or concavity. Hanyga and Slawinski (2001) studied caustics and qSV ray fields of transversely isotropic and vertically inhomogeneous media. Waheed and Alkhalifah (2016) suggested an effective ellipsoidal model for tilted orthorhombic (TOR) media, where the computed wavefield contains most of the critical components, including frequency dependency and caustics. Červený and Pšenčík (2017) applied a paraxial Gaussian beam approach to study wavefields at caustics and their vicinities, computing both geometric spreading and phase shifts in inhomogeneous anisotropic complex media. Mitrofanov and Priimenko (2018) applied the KMAH index to separate compressional and converted waves, in particular PSP waves from streamer data. A comprehensive discussion on caustics is given in the book by Arnold (1994), with an extensive list of references therein.

Traditionally in seismology, the KMAH index is obtained by solving the DRT equation set for matrix $\mathbf{Q}$ and its accompanying matrix $\mathbf{P}$,

$$Q_{ij} = \frac{\partial x_i}{d\gamma_j} \quad , \quad P_{ij} = \frac{\partial p_i}{\partial \gamma_j} = \frac{\partial^2 t}{\partial x_i \partial \gamma_j} \quad , \quad i, j = 1, 2, 3, \quad \gamma_3 \equiv \tau \qquad , \tag{1}$$

where $p_i$ are the slowness vector components, $\gamma_j$ are the point source ray coordinates (RC) with , for example, $\gamma_3 \equiv \tau$, the current traveltime along the central ray (other ray characteristics for $\gamma_3$ can be used as well), and $x_i$ are the Cartesian coordinates. The matrix product $\mathbf{M}$,

$$\mathbf{M} = \mathbf{P}\mathbf{Q}^{-1} = \frac{d^2 t}{d\mathbf{x}^2} = \frac{d\mathbf{p}}{d\mathbf{x}} \qquad , \tag{2}$$

(e.g., Červený, 2000) represents the spatial gradient of the slowness vector along the ray. A further discussion about the physical characteristics of matrix **M** is given in Appendix C of Part VI. The conditions for the types of caustics can be derived from the transformation matrix **Q,** analyzing the zeros of its determinant. It is, however, convenient to rotate the matrix from the global to a local Cartesian frame where the third axis is tangent to the ray, and to reduce the dimension of **Q** from 3 × 3 to 2 × 2, $\mathbf{Q} \rightarrow \mathbf{Q}_{2\mathrm{x}2}$ . For the case where $\gamma_3 = \tau$ (the current time), the third column of the matrix $\mathbf{Q}$, $\partial \mathbf{x} / \partial \gamma_3 = \partial \mathbf{x} / \partial \tau = \mathbf{v}_{\mathrm{ray}}$ includes the ray (group) velocity components. After the rotation, two components of this column are zeros and the third is the absolute value of the ray velocity, $v_{\mathrm{ray}}$. The Jacobian $J^{(\tau)}$ of the ray tube along the central ray is given by the determinant $J^{(\tau)} = \det \mathbf{Q} = v_{\mathrm{ray}} \det \mathbf{Q}_{2\mathrm{x}2}$, where $\mathbf{Q}_{2\mathrm{x}2}$ is the upper left 2 x 2 block of the 3 x 3 matrix $\mathbf{Q}$ in the local frame. Note that $\det \mathbf{Q}$ has units of a flow rate volume over time, [$L^3/T$], while $\det \mathbf{Q}_{2x2}$ has units of area, [$L^2$]. Note also that the rotation to the local Cartesian frame does not affect $\det \mathbf{Q}$. It is important to note that the units of the ray Jacobian depend on the flow parameters used in its computation. In this study, we use the arclength $s$ as the flow parameter, and the ray Jacobian $J \equiv J^{(s)} = J^{(\tau)} / v_{\text{ray}}$ has units of area, [$L^2$] . Hence, in order to identify and characterize possible caustics, it is sufficient to analyze $\mathbf{Q}_{2x2}$. If only its determinant vanishes (a single zero eigenvalue) , we deal with a first-order (line) caustic, leading to an additional phase shift of $\pi / 2$. If both the determinant and the trace of the matrix vanish (both eigenvalues are zero), this is a second-order (point) caustic, leading to an additional phase shift of $\pi$. The cumulative number of $\pi / 2$ phase shifts, $\kappa_J$, along the ray is the KMAH index. The complex valued amplitude of the ray (e.g., the particle displacement) is related to the ray Jacobian (which can also be negative) under the square root. It can be presented as a number with a magnitude and phase, where the latter is defined by the KMAH index,

$$\sqrt{J} = \sqrt{|J|} \ \exp\left(-\frac{i\pi\kappa_J}{2}\right) \qquad . \qquad (3)$$

The minus sign of the phase shift, due to caustic, is chosen in order to keep the sign convention of the analytical signal (a complex-valued function with no negative frequency components),

$$F(t) = f(t) + i\,\text{HT}[\,f(t)] \qquad , \qquad (4)$$

where $f(t)$ is the source signal and HT is its Hilbert transform (e.g., Červený, 2000; Schleicher et al., 2007). For isotropic media and for compressional waves in anisotropic media, the occurrence of each caustic increases the KMAH index by 1 or 2, depending on the caustic order. For shear waves in anisotropic media, the change of the KMAH index at a caustic can also accept negative values $-1$ and $-2$, since the slowness surface for these waves is not necessarily convex (Bakker, 1998; Červený, 2000; Klimeš, 2010, 2014).

In our study, we work with both, local Cartesian coordinates closely related to the RCC, where the flow parameter is oriented in the direction of the ray (i.e., the ray velocity direction), and with the global Cartesian frame. The local Cartesian coordinate systems can also be defined for a flow parameter oriented in the direction of the slowness vector, to be related to the wavefront orthonormal coordinates (WOC) (Červený, 2000).

We note that in this work we do not deal with asymptotic high-frequency methods, such as the Gaussian beam or Maslov, primarily designed to overcome the (singularity) problems of the standard ray theory in the computation of dynamic properties across caustics. Indeed, the Gaussian-beam summation method (Červený et al., 1982; Popov, 1982; Červený, 1985; Nowack, 2003, among others), where the beam packets are expressed in terms of complex-valued vector amplitudes and complex-valued traveltime (phase), has been widely used in the seismic method. However, at this stage, this method is beyond the scope of our study.

## NORMALIZED GEOMETRIC SPREADING

The geometric spreading $L_{GS}$ is directly related to the ray Jacobian. It has units $[\mathrm{L}^2/\mathrm{T}]$, where L stands for distance and T for time. The units of the ray Jacobian depend on the parameters used in its computation. In this study, the flow parameter is the arclength and the ray Jacobian $J \equiv J^{(s)}$ has units of area, $[\mathrm{L}^2]$. In cases where the flow parameter is the traveltime, the relation between the Jacobians is $J^{(\tau)} = J\, v_{\mathrm{ray}}$ (e.g., Červený, 2000). Both Jacobians represent the flow rate. $J^{(\tau)}$ is a volume per unit time, while $J$ is a volume per unit length. In cases where the flow parameter is $\sigma$ , defined in equation 5 below, the relation is $J^{(\sigma)} = J / v_{\text{ray}}$. Unlike geometric spreading, which is a physical reciprocal characteristic of the ray, the value of the Jacobian depends on the "direction" of the DRT computation (from endpoint $S$ to endpoint $R$, with the point-source initial conditions at $S$ or vice versa, with the initial conditions at $R$). In Appendix F, we describe the relationship between the ray Jacobian and the geometric spreading for general anisotropy using our proposed ray coordinates (RC).

The geometric spreading $L_{GS}$ can be normalized by the parameter $\sigma$ that has the same units $[\mathrm{L}^2/\mathrm{T}]$, and can be defined (and computed) along the ray path as,

$$\sigma = \int_S^R v_{\text{ray}}^2(\tau)\, d\tau = \int_S^R v_{\text{ray}}(s)\, ds \qquad , \qquad (5)$$

where $\tau$ and $s$ are the running (flow) parameters: the time and arclength along the path, respectively. For isotropic models with a constant, not necessarily vertical, velocity gradient, $L_{GS} = \sigma$. We define the ratio $L_{GS} / \sigma$ as the normalized geometric spreading; , it provides a certain level "feeling" (intuition) about the effect of the anisotropy and the change of the velocity gradient on the actual value of the geometric spreading. In some seismic modeling/imaging applications, parameter $\sigma$ is used to approximate the actual geometric spreading, $L_{GS} \approx \sigma$, but we explicitly show in our numerical examples that even for simple 1D models, with a monotonously (but non-uniformly) increasing velocity with depth, the error of such a substitution may be significant, and for more complex models this approximation is wrong. Parameter $\sigma$ can only increase along the ray, while the geometric spreading may increase and decrease, and it vanishes at caustic points. At the source point, for isotropic media, $L_{GS} / \sigma = 1$; along the ray, this ratio may be below or above 1. For anisotropic media, this ratio differs from 1 even at the source point (see Part VI for details and derivation).

## TYPES OF STATIONARY PATHS AND EXISTENCE OF CAUSTICS

For the stationary ray path resulting from the kinematic analysis, we first compute the eigenvalues of its global traveltime Hessian matrix to check whether the path delivers a minimum traveltime solution. If all eigenvalues are positive, the stationary path is a minimum traveltime. We assume that the (hypothetic) case where all the eigenvalues are negative (a maximum traveltime solution) is not realistic for systems with multiple DoF. If one or more eigenvalues are negative, and the others are positive, it is a saddle point solution, which is an indication of the existence of caustics: zeros of the (signed) cross-section area of the ray tube (ray Jacobian) along the ray. We note that this type of eigenvalue analysis of the global traveltime Hessian cannot provide the actual number of caustics, their locations and type: line or point, which are important for implementing the required phase correction of the Green's function. This information requires explicit solution of the dynamic ray tracing vector equation, which is the topic of Parts V, VI and VII of the Eigenray study.

Following the theory presented in Part I, the traveltime stationary condition for a ray path between two endpoints, $S$ and $R$, is given by,

$$t = \int_S^R L\left[\mathbf{x}(s), \mathbf{r}(s)\right] ds \rightarrow \text{stationary} \ , \quad \mathbf{r} = \dot{\mathbf{x}} = d\mathbf{x} / ds \ , \quad \mathbf{r} \cdot \mathbf{r} = 1 \quad , \tag{6}$$

where the flow parameter is the arclength $s$, and the integrand (Lagrangian) $L$ is formulated as,

$$L(\mathbf{x},\mathbf{r}) = \frac{dt}{ds} = \frac{\sqrt{\mathbf{r}\cdot\mathbf{r}}}{v_{\mathrm{ray}}(\mathbf{x},\mathbf{r})} = \frac{\sqrt{\mathbf{r}\cdot\mathbf{r}}}{v_{\mathrm{ray}}[\mathbf{C}(\mathbf{x}),\mathbf{r}]} \quad . \tag{7}$$

Here, $v_{\mathrm{ray}}$ is the ray (group) velocity, and $\mathbf{C}(\mathbf{x})$ is the density-normalized symmetric stiffness matrix of dimension 6. The invariance of the traveltime integral in equation 6, under a change of the flow parameter $s$, means that the Lagrangian $L$ (equation 7) complies with the first-order homogeneity assumption,

$$L(\mathbf{x},k\,\mathbf{r}) = k\,L(\mathbf{x},\mathbf{r}) \quad , \tag{8}$$

wrt the ray direction $\mathbf{r}$ (Bliss, 1916), where $k$ is any positive number. Note that other flow parameters, like the current time $\tau$ or parameter $\sigma$ can be used as well. Several important properties of a parametric function follow from this linear homogeneity. In particular, due to Euler's theorem,

$$L_{\mathbf{r}}\cdot\mathbf{r} = L \quad (\text{indeed,} \quad L_{\mathbf{r}}\cdot\mathbf{r} = \mathbf{p}\cdot\mathbf{r} = v_{\mathrm{ray}}^{-1} = L) \quad , \tag{9}$$

where vectors $L_{\mathbf{x}}(s)$ and $L_{\mathbf{r}}(s)$ of length 3 are the location and directional gradients of the Lagrangian. They are derived as explicit functions of the ray velocity, $v_{\mathrm{ray}}$, and its spatial and directional gradients, $\nabla_{\mathbf{x}} v_{\mathrm{ray}}$ and $\nabla_{\mathbf{r}} v_{\mathrm{ray}}$, in Appendix A. Derivatives of equation 9 wrt the location $\mathbf{x}$ and direction $\mathbf{r}$ provide the following interesting relations (Bliss, 1916),

$$L_{\mathbf{xr}}\cdot\mathbf{r} = \mathbf{r}\cdot L_{\mathbf{rx}} = L_{\mathbf{x}} \quad , \tag{10}$$

and,

$$L_{\mathbf{rr}} \cdot \mathbf{r} = 0 \qquad . \tag{11}$$

Equation 11 is also an evidence that the ray direction $\mathbf{r}$ is an eigenvector of the positive semidefinite matrix $L_{\mathbf{rr}}$, and the corresponding eigenvalue is zero. According to variational calculus (e.g., Gelfand and Fomin, 2000), the positive semidefinite matrix $L_{\mathbf{rr}}$ is a local criterion of the necessary (but not sufficient) conditions for a minimum. The sufficient conditions for a minimum include local and global counterparts. The local sufficient condition requires for $L_{\mathbf{rr}}$ to be a positive definite matrix (which is not the case). The global sufficient condition for a minimum is the absence of caustics. This global condition is also a necessary one.

However, practically, the global criterion of both necessary and sufficient conditions for a minimum traveltime is the absence of caustics, which can be established by solving the proposed Jacobi DRT equation. In Appendix B we further discuss and simplify the necessary and sufficient conditions for a minimum, introducing additional assumptions.

## THE PROPOSED JACOBI DRT EQUATION SET

Given a stationary ray path (referred as a central ray) $\mathbf{x}(s)$, with proper initial conditions of its corresponding ray tube (described in the next section), the proposed Jacobi DRT equation enables to characterize the geometry of the ray tube around the central ray, in particular, the shift (displacement) vector, $\mathbf{u}(s) = \delta\mathbf{x}(s)$, of a paraxial ray, $\mathbf{x}_{\text{prx}}(s) = \mathbf{x}(s) + \mathbf{u}(s)$. At each point along the arclength $s$ of the central ray, the shift $\mathbf{u}(s)$ belongs to a plane normal to the central ray direction, $\mathbf{r}(s)$. A scheme of a ray tube for point-source paraxial rays is presented in Figure 1.

For point-source initial conditions, a set of two normal shift vectors, $\mathbf{u}_1(s)$ and $\mathbf{u}_2(s)$, is required to further compute the ray Jacobian $J(s)$ in order to establish the geometric spreading $L_{GS}(s)$ along the ray path, to identify possible caustic locations and to classify the caustic type (order): first order (line) and second order (point), i.e., to compute the KMAH index $k_J(s)$.

<u>The Jacobi DRT Equation</u>

Recall that in Part I (equation 8) we obtained the KRT equation by applying the Euler-Lagrange equation, $dL_{\mathbf{r}} / ds = L_{\mathbf{x}}$ , to the Lagrangian (traveltime integrand), which is the condition of the vanishing first traveltime variation. In this study, we further follow Bliss (1916) and apply the Euler-Lagrange equation to the second traveltime variation (Appendix C). This leads to the vector-form, linear, second-order, Jacobi DRT equation,

$$\frac{d}{ds}\left(L_{\mathbf{rx}} \cdot \mathbf{u} + L_{\mathbf{rr}} \cdot \dot{\mathbf{u}}\right) = L_{\mathbf{xx}} \cdot \mathbf{u} + L_{\mathbf{xr}} \cdot \dot{\mathbf{u}} \quad , \tag{12}$$

where (as already mentioned) vector $\mathbf{u}(s) = \delta\mathbf{x}(s)$ is a vector solution of this set, representing a normal shift between a paraxial ray $\mathbf{x}_{\text{prx}}(s)$ and the central ray $\mathbf{x}(s)$, and $\dot{\mathbf{u}}(s) = \delta\mathbf{r}(s)$ is the corresponding direction change. Normal shift means that vector $\mathbf{u}$ belongs to the plane normal to the ray direction $\mathbf{r}$,

$$\mathbf{u} \cdot \mathbf{r} = 0 \quad \text{and} \quad \frac{d}{ds}\left(\mathbf{u} \cdot \mathbf{r}\right) = \dot{\mathbf{u}} \cdot \mathbf{r} + \mathbf{u} \cdot \dot{\mathbf{r}} = 0 \quad . \tag{13}$$

In the next section we show that, depending on the initial conditions, the normal shift $\mathbf{u}$ for a general paraxial ray consists of a linear combination of (up to four) basic normal shift solutions $\mathbf{u}_i$ that differ by their initial conditions.

Note that the Jacobi DRT equation (equation12) includes a second derivative of the normal shift wrt the arclength, $\dfrac{d^2\mathbf{u}}{ds^2} = \ddot{\mathbf{u}}(s) = \dfrac{d\dot{\mathbf{u}}}{ds}$. In Part VII, we show how this second derivative can be eliminated using the finite-element approach with the weak formulation. The detailed derivation of the Jacobi DRT equation (equation 12) is presented in Appendix C.

The Jacobi DRT equation is a linear ODE with varying coefficients and includes four arclength-dependent, second-order tensors, representing the second derivatives of the Lagrangian $L(s) = dt/ds$, wrt the Cartesian components of the ray location and direction: $L_{\mathbf{xx}}(s), L_{\mathbf{rr}}(s), L_{\mathbf{xr}}(s), L_{\mathbf{rx}}(s)$. (i.e., a spatial, a directional, and two mutually transposed mixed tensors). These tensors have been derived in Appendix A. In Appendix B, we study the necessary and sufficient conditions for a minimum traveltime stationary ray path. This is mainly important for the identification of possible caustics. As mentioned, the information about the existence of caustics can be obtained prior to the actual solution of the DRT equation by the eigenvalue analysis of the global traveltime Hessian.

Direction of a paraxial ray

Note that $\mathbf{r}(s)$ is the direction of the central ray $\mathbf{x}(s)$ and not of the paraxial ray, $\mathbf{x}_{\text{prx}}(s) = \mathbf{x}(s) + \mathbf{u}(s)$. Also, the flow parameter $s$ is the arclength of the central ray and not of the paraxial ray. In order to compute the direction of the paraxial ray wrt the arclength parameter of the central ray, $\mathbf{r}_{\text{prx}}(s)$, we differentiate the paraxial ray path wrt parameter *s*, $\dot{\mathbf{x}}_{\text{prx}}(s)=\dot{\mathbf{x}}(s)+\dot{\mathbf{u}}(s)=\mathbf{r}(s)+\dot{\mathbf{u}}(s)$ , and then normalize it to obtain,

$$\mathbf{r}_{\text{prx}}(s)=\frac{\dot{\mathbf{x}}_{\text{prx}}(s)}{\left|\dot{\mathbf{x}}_{\text{prx}}(s)\right|}=\frac{\mathbf{r}(s)+\dot{\mathbf{u}}(s)}{\sqrt{\left[\mathbf{r}(s)+\dot{\mathbf{u}}(s)\right]\cdot\left[\mathbf{r}(s)+\dot{\mathbf{u}}(s)\right]}}=\frac{\mathbf{r}(s)+\dot{\mathbf{u}}(s)}{\sqrt{\mathbf{r}^2(s)+2\mathbf{r}(s)\cdot\dot{\mathbf{u}}(s)+\dot{\mathbf{u}}^2(s)}} \quad . \tag{14}$$

Since $\mathbf{r}^2=1$, it becomes,

$$\mathbf{r}_{\text{prx}}(s)=\mathbf{r}(s)+\dot{\mathbf{u}}(s)-\mathbf{r}(s)\left[\mathbf{r}(s)\cdot\dot{\mathbf{u}}(s)\right]+O\left(\dot{\mathbf{u}}^2\right) \quad . \tag{15}$$

Taking into account the constraint of equation 13, the paraxial direction can be rearranged as,

$$\mathbf{r}_{\text{prx}}(s)=\mathbf{r}(s)+\dot{\mathbf{u}}(s)+\mathbf{r}(s)\left[\dot{\mathbf{r}}(s)\cdot\mathbf{u}(s)\right]+O\left(\dot{\mathbf{u}}^2\right) \quad , \tag{16}$$

where $\dot{\mathbf{r}}(s)$ is the vector-form curvature of the central ray. Note that the direction difference vector $\mathbf{r}_{\text{prx}}(s)-\mathbf{r}(s)$ is normal to the direction of the central ray $\mathbf{r}(s)$, ignoring the second-order terms, which leads to, $\left[\mathbf{r}_{\text{prx}}(s)-\mathbf{r}(s)\right]\cdot\mathbf{r}(s)=O\left(\dot{\mathbf{u}}^2\right)$.

## INITIAL CONDITIONS FOR A GENERAL PARAXIAL RAY

Different initial conditions (IC) or boundary conditions (BC) (or even mixed conditions) can be set, in order to solve the proposed Jacobi DRT equation 12 and define paraxial rays. In this section we focus on the of initial conditions.

The vector-form Jacobi DRT equation 12 is actually a set of three linear second-order scalar equations. Generally, in similar ODE cases, each Cartesian component of the unknown vector requires two initial conditions (e.g., values of the component and its derivative). However, due to the first-degree homogeneity of the parametric Lagrangian $L(s)$ wrt the ray direction $\mathbf{r}$, only two (of the three) scalar second-order equations are independent (see Appendix C for a detailed proof of this property). Therefore, the most general normal shift vector (defining a paraxial ray) that represents a fundamental solution $\mathbf{u}$ of the Jacobi DRT equation, includes only four independent basic solutions $\mathbf{u}_i(s)$, $i=1,...,4$, corresponding to the four different initial conditions (IC) $\mathbf{u}_i(S)$, $i=1,...,4$ (where $S$ indicates the source location, $s=0$), and ray coordinates (RC), $\gamma_j$, $j=1,...,4$. The Jacobi DRT equation is linear, and therefore, its fundamental solution $\mathbf{u}$ is a linear combination of the four basic solutions $\mathbf{u}_i$, where the scalar RC $\gamma_i$ are the corresponding weights,

$$\mathbf{u}=\sum_{i=1}^{4}\gamma_i\,\mathbf{u}_i \quad . \tag{17}$$

In this study, $\gamma_1$ and $\gamma_2$ represent point-source RC associated with the corresponding basic solutions $\mathbf{u}_1(s)$ and $\mathbf{u}_2(s)$, and $\gamma_3$ and $\gamma_4$ represent plane-wave RC associated with the corresponding basic solutions $\mathbf{u}_3(s)$ and $\mathbf{u}_4(s)$. The general paraxial ray can be written as a linear combination of these two pairs of basic solutions. The basic solution vectors are normal to the direction of the central ray along the whole path; they represent the normal shifts between the paraxial ray and the central ray vs. the arclength of the central ray. Furthermore, a specific paraxial ray can be defined by arbitrary valid IC or BC, that do not coincide with the IC of any basic solution. Such ray can be still presented as a linear combination of the basic solutions, with the four RC $\gamma_i$ established from the IC/BC of the given paraxial ray.

Note that in this study, we work with the Cartesian components of the normal shifts $\mathbf{u}_i$. However, since this vector is in the plane normal to the ray, it can be rotated to a local Cartesian frame, $x_3^{\text{loc}} = \mathbf{r}$, where it can be defined by two components only, in the plane normal to the ray (similar in a sense to the RCC).

## INITIAL CONDITIONS FOR A POINT-SOURCE PARAXIAL RAY

In this study we are primarily interested in the computation of geometric spreading and identification of caustics along the ray (the amplitude and phase shifts of the asymptotic Green function). For this, only the two basic solutions, $\mathbf{u}_1$ and $\mathbf{u}_2$, and their corresponding RC coefficients $\gamma_1$ and $\gamma_2$, related to point-source paraxial rays, should be studied. At the start point $S$, $\mathbf{u}_{1,S} \equiv \mathbf{u}_1(S)$ and $\mathbf{u}_{2,S} \equiv \mathbf{u}_2(S)$ vanish. The IC for the point-source basic solutions are constructed in Appendix D and can be summarized as,

$$\begin{array}{lll} L_{\mathbf{rr},S}\dot{\mathbf{u}}_{1,S} = \lambda_{1,S}\dot{\mathbf{u}}_{1,S} \ , & L_{\mathbf{rr},S}\dot{\mathbf{u}}_{2,S} = \lambda_{2,S}\dot{\mathbf{u}}_{2,S} \ , & L_{\mathbf{rr},S}\mathbf{r}_S = 0 \ , \\ \mathbf{u}_{1,S} = 0 \ , & \mathbf{u}_{2,S} = 0 \ , & \dot{\mathbf{u}}_{1,S} \times \dot{\mathbf{u}}_{2,S} \cdot \mathbf{r}_S = 1 \ . \end{array} \tag{18}$$

Obviously the normal vector shifts at the source, $\mathbf{u}_{1,S}$ and $\mathbf{u}_{2,S}$, vanish; their arclength derivatives, $\dot{\mathbf{u}}_{1,S}$ and $\dot{\mathbf{u}}_{2,S}$, become the eigenvectors of matrix $L_{\mathbf{rr},S}$. For anisotropic media, these eigenvectors correspond to the distinct nonzero eigenvalues $\lambda_{1,S}$ and $\lambda_{2,S}$, normalized to the unit length, $|\dot{\mathbf{u}}_{1,S}| = |\dot{\mathbf{u}}_{2,S}| = 1$.

In isotropic media, the two nonzero (positive) eigenvalues $\lambda_{1,S}$ and $\lambda_{2,S}$ coincide, and the two eigenvectors are not fully defined; we only know that $\dot{\mathbf{u}}_{1,S}$ and $\dot{\mathbf{u}}_{2,S}$ are both normal to the ray direction $\mathbf{r}_S$ (which follows from equation 13 when we set $\mathbf{u}_{i,S} = 0$) and to each other. In order to fully define these vectors, we set $\dot{\mathbf{u}}_{1,S}$ to be in the plane of the central ray path and $\dot{\mathbf{u}}_{2,S}$ in its normal plane.

Hence, the general point-source paraxial ray $\mathbf{x}_{\text{prx}}(\gamma_1, \gamma_2, s)$ represents a linear combination of the central ray $\mathbf{x}(s)$ and the two basic solutions, $\mathbf{u}_1(s)$ and $\mathbf{u}_2(s)$,

$$\mathbf{x}_{\text{prx}}(\gamma_1, \gamma_2, s) = \mathbf{x}(s) + \gamma_1 \mathbf{u}_1(s) + \gamma_2 \mathbf{u}_2(s) \quad , \tag{19}$$

where $\gamma_1$ and $\gamma_2$ are the (infinitesimal) point source ray coordinates (RC), defined as the coefficients of the basic normal shift vector solutions $\mathbf{u}_1$ and $\mathbf{u}_2$ that constitute the fundamental point-source solution. The RC, $\gamma_1$ and $\gamma_2$, do not vanish simultaneously, $\gamma_1^2 + \gamma_2^2 \neq 0$.

Note that at the end of Appendix E, the RC coefficients $\gamma_1$ and $\gamma_2$ are explicitly derived for a point source with a given vector shift $\Delta \mathbf{x}_R$ at the endpoint (the receiver).

## INITIAL CONDITIONS FOR A PLANE-WAVE PARAXIAL RAY

Disclaimer: A plane-wave paraxial ray is normally defined by its slowness vector, which is collinear to the slowness vector of the central ray at the start point $S$. This is not the case in our study, where the plane-wave paraxial rays share the same ray direction as the ray direction of the central ray at $S$. This definition naturally follows from the defined DoF of the Jacobi KRT and DRT equation; these DoF consist of the ray locations and ray directions.

The plane-wave RC $\gamma_3$ and $\gamma_4$ are defined as the coefficients of the basic normal shifts $\mathbf{u}_3(s)$ and $\mathbf{u}_4(s)$ constituting the fundamental plane-wave solution. At the source, the paraxial ray $\mathbf{x}_{\mathrm{prx},S}$ is collinear with the central ray $\mathbf{x}_S$. The basic solutions at the source, $\mathbf{u}_{3,S}$ and $\mathbf{u}_{4,S}$, are normal to the central ray and to each other, and are normalized to the unit length. The details for construction of the plane-wave paraxial rays' IC are given in Appendix D. We assign the normalized eigenvectors of matrix $L_{\mathbf{rr},S}$ as the start-point normal shifts $\mathbf{u}_{3,S}$ and $\mathbf{u}_{4,S}$,

$$
\begin{array}{lll}
L_{\mathbf{rr},S}\mathbf{u}_{3,S} = \lambda_{1,S}\mathbf{u}_{3,S} \quad , & L_{\mathbf{rr},S}\mathbf{u}_{4,S} = \lambda_{2,S}\mathbf{u}_{4,S} \quad , & L_{\mathbf{rr},S}\mathbf{r}_S = 0 \quad , \\
\dot{\mathbf{u}}_{3,S} = -\left(\dot{\mathbf{r}}_S \cdot \mathbf{u}_{3,S}\right)\mathbf{r}_S \quad , & \dot{\mathbf{u}}_{4,S} = -\left(\dot{\mathbf{r}}_S \cdot \mathbf{u}_{4,S}\right)\mathbf{r}_S \quad , & \mathbf{u}_{3,S} \times \mathbf{u}_{4,S} \cdot \mathbf{r}_S = 1 \quad .
\end{array}
\tag{20}
$$

Similarly to the point-source paraxial ray, the case of the multiple (double) nonzero eigenvalues, $\lambda_{1,S} = \lambda_{2,S}$, requires a special treatment (see details in Appendix D). The general plane-wave paraxial ray $\mathbf{y}(\gamma_1, \gamma_2, s)$ represents a linear combination of the central ray $\mathbf{x}(s)$ and the two basic solutions, $\mathbf{u}_3(s)$ and $\mathbf{u}_4(s)$,

$$
\mathbf{x}_{\mathrm{prx}}(\gamma_3, \gamma_4, s) = \mathbf{x}(s) + \gamma_3 \mathbf{u}_3(s) + \gamma_4 \mathbf{u}_4(s) \qquad , \tag{21}
$$

where $\gamma_3$ and $\gamma_4$ are the (infinitesimal) plane-wave RC, and $\gamma_3^2 + \gamma_4^2 \neq 0$.

## A GENERAL PARAXIAL RAY

To obtain a general paraxial ray, one can combine solutions of equations 19 and 21, yielding the fundamental solution of the Jacobi DRT equation,

$$\underbrace{\mathbf{x}_{\text{prx}}\left(\gamma_1,\gamma_2,\gamma_3,\gamma_4,s\right)}_{\text{paraxial ray}} = \underbrace{\mathbf{x}\left(s\right)}_{\text{central ray}} + \underbrace{\gamma_1\mathbf{u}_1\left(s\right)+\gamma_2\mathbf{u}_2\left(s\right)}_{\text{point source}} + \underbrace{\gamma_3\mathbf{u}_3\left(s\right)+\gamma_4\mathbf{u}_4\left(s\right)}_{\text{plane wave}} \quad . \tag{22}$$

The four basic solutions, $\mathbf{u}_i,\ i=1,\ldots 4$, are defined by their corresponding initial conditions. Thus, the four RC, $\gamma_i,\ i=1,\ldots 4$ are also fully defined for any paraxial ray. The Jacobi DRT equation allows the application of alternative initial conditions (IC) or boundary conditions (BC) that do not coincide with the IC of any basic solution, as well as also mixed IC or BC (involving both the function and its derivative in a single condition) when, for example, the ratio between the function and its derivative is specified. However, this does not lead to new solutions, as any paraxial ray can be presented by the fundamental solution of equation 22. Such ray can be still presented as a linear combination of the basic solutions, with the four RC $\gamma_i$ established from the IC/BC of the given paraxial ray.

The combined point-source and plane-wave initial conditions are normally used to define the so-called propagator matrix. However, in this study we are primarily interested in computation of the point-source dynamic parameters, which are needed to establish geometric spreading and to detect possible caustics.

Note that due to the linearity of the Jacobi DRT equation, should we apply a different nonzero absolute value of the normal shift derivative at the source point, $\left|\dot{\mathbf{u}}_{i,S}\right| \neq 1,\ i=1,2$ (or to the normal shift itself in the case of a plane wave, $\left|\mathbf{u}_{i,S}\right| \neq 1,\ i=3,4$), the solution increases or decreases proportionally to this initial value. That is why a paraxial solution $\mathbf{x}_{\text{prx}}(s)$ depends linearly on the RC $\gamma_i$ and on the basic solutions $\mathbf{u}_i$ in equations 19, 21 and 22.

## RAY JACOBIAN, GEOMETRIC SPREADING AND CAUSTIC CRITERIA

As mentioned, the ray Jacobian can be viewed as the determinant of the transform matrix $\mathbf{Q}$ between the Cartesian coordinates and the RC, for a point source paraxial ray, $\mathbf{x}_{\text{prx}}(\gamma_1,\gamma_2,s)$, computed at a point of the central ray, $\gamma_1=\gamma_2=0$,

$$J(s)=\frac{\partial\mathbf{x}_{\text{prx}}}{\partial\gamma_1}\times\frac{\partial\mathbf{x}_{\text{prx}}}{\partial\gamma_2}\cdot\frac{\partial\mathbf{x}_{\text{prx}}}{\partial s}=\det\underbrace{\begin{bmatrix}\dfrac{\partial\mathbf{x}_{\text{prx}}}{\partial\gamma_1} & \dfrac{\partial\mathbf{x}_{\text{prx}}}{\partial\gamma_2} & \dfrac{\partial\mathbf{x}_{\text{prx}}}{\partial s}\end{bmatrix}}_{\text{columns}}=\det\mathbf{Q} \quad , \tag{23}$$

where the first two ray coordinates, $\gamma_1$ and $\gamma_2$, appear in point-source equation 19, and the third is the arclength $s$. This leads to,

$$J(s)=\mathbf{u}_1(s)\times\mathbf{u}_2(s)\cdot\mathbf{r}(s) \quad . \tag{24}$$

The Jacobian represents a signed cross-area of the ray tube per unit RC, and the vanishing Jacobian is an evidence of the intersection of a paraxial ray $\mathbf{x}_{\text{prx}}(s)$ with the central ray $\mathbf{x}(s)$, which in turn, means a caustic.

The difference between the locations of the point-source paraxial ray and the central ray reads,

$$\mathbf{x}_{\text{prx}}(\gamma_1, \gamma_2, s+\Delta s) - \mathbf{x}(s) = \mathbf{Q}\begin{bmatrix}\gamma_1\\ \gamma_2\\ \Delta s\end{bmatrix} = [\mathbf{u}_1 \quad \mathbf{u}_2 \quad \mathbf{r}]\cdot\begin{bmatrix}\gamma_1\\ \gamma_2\\ \Delta s\end{bmatrix} \qquad , \tag{25}$$

where $\Delta s$ is an infinitesimal increment of the arclength along the central ray.

<u>Geometric Spreading</u>

Geometric spreading can be computed as suggested, for example, by Červený (2000) (equation 4.14.46) . In our notations this relationship looks like,

$$L_{GS}(s) = v_{J,S}\sqrt{\frac{v_{\text{ray}}(s)}{v_{\text{phs}}(s)}|\, J(s)|} \qquad , \tag{26}$$

where $v_{\text{phs}}(s)$ and $v_{\text{ray}}(s)$ are the phase and ray velocities, respectively, at any current point along the arclength $s$ of the ray path. Parameter $v_{J,S}$ in equation 26 is considered a "conversion velocity" factor that only depends on the point-source initial conditions; its computation is explained in Appendix F. As already mentioned, the geometric spreading has units $\left[\mathrm{L}^2/\mathrm{T}\right]$, where $\mathrm{L}$ stands for distance and $\mathrm{T}$ for time. The units of the ray Jacobian depend on the parameters used in its computation. In this study, the flow parameter is the arclength and the ray Jacobian $J \equiv J^{(s)}$ has units of area, $\left[\mathrm{L}^2\right]$. Note that in our notation, any parameter indicated with a subscript uppercase $S$ refers to the point source, and lowercase $s$ - to the arclength.

For isotropic media, with the RC type chosen in this study, this factor simplifies to, $v_{J,S} = v_S$, where $v_S$ is the medium velocity at the source,

$$L_{GS}^{\text{iso}}(s) = v_S \sqrt{|J(s)|} \quad . \tag{27}$$

For a general anisotropic case, we obtain the following expression for $v_{J,S}$,

$$v_{J,S} = \frac{1}{\sqrt{\lambda_{1,S}\lambda_{2,S}}} \sqrt{\frac{v_{\text{ray},S}}{v_{\text{phs},S}}} \quad , \tag{28}$$

where $\lambda_{1,S}$ and $\lambda_{2,S}$ (see equation 18) are the nonzero eigenvalues of matrix $L_{\mathbf{rr},S}$ (normally positive), and $v_{\text{phs},S}$ and $v_{\text{ray},S}$ are the phase and ray velocities, respectively, at the source. Using equation 28, we assume that the initial conditions of the Jacobi DRT equation for the two ray coordinates (RC), $\dot{\mathbf{u}}_{1,S}$ and $\dot{\mathbf{u}}_{2,S}$, are the (normalized) eigenvectors of matrix $L_{\mathbf{rr},S}$, normal to the ray and to each other, corresponding to its nonzero eigenvalues. The third eigenvalue of matrix $L_{\mathbf{rr}}$ is zero, and the corresponding eigenvector is the ray direction $\mathbf{r}$. The eigenvalues have the units of slowness. In an isotropic case, $\lambda_{1,S} = \lambda_{2,S} = v_S^{-1}$, where $v_S$ is the medium velocity at the source. The initial conditions represent the derivatives of the normal paraxial shifts wrt the arclength of the central ray at the source. Note that $v_{J,S}$ is not a physical velocity, but an appropriate conversion coefficient with the units of velocity. Combining equations 26 and 28, we obtain the relationship between the geometric spreading and the ray Jacobian,

$$L_{GS}(s) = \sqrt{\frac{v_{\text{ray},S}}{v_{\text{phs},S}} \frac{v_{\text{ray}}(s)}{v_{\text{phs}}(s)} \frac{|J(s)|}{\lambda_{1,S}\lambda_{2,S}}} \quad . \tag{29}$$

We re-emphasize that the ray Jacobian depends on the chosen RC, while the geometric spreading does not - it is a fundamental physical parameter of the wave/ray propagation. Note that the simple form of equation 29 is only valid for the RC type proposed in this paper (see equation 18). The general form is given by Červený (2000) (equation 4.14.46), and includes the determinant of the matrix whose columns are the slowness vector derivatives wrt the RC (see Appendix F, equation F6). The matrix is computed at the source, in the wavefront orthonormal coordinates (WOC) reference frame, and then reduced from dimension $3\times3$ to $2\times2$ by removing the row and column related to the slowness direction.

In Appendix F we derive and explain the computation of the ray Jacobian, the criteria for the two types of caustics: first-order (line) and second-order (point), a way to define the direction of a line caustic, and the relationship between the ray Jacobian and geometric spreading in general anisotropic media (equation 29).

## RAY PATH COMPLEXITY CRITERION

In order to estimate the reliability/plausibility of the ray path obtained with the Eigenray KRT, we introduce the (weighted) propagation complexity criterion, similar to the endpoint complexity criterion introduced in equation 10 of Part IV, but now accounting for the normalized relative geometric spreading along the whole ray rather than at the destination point only,

$$c_r = \frac{1}{s_{\max}} \int_S^R \left[ \frac{L_{GS}(s)}{\sigma(s)} - 1 \right]^2 ds \qquad , \qquad (30)$$

where $s_{\max}$ is the full arclength of the ray path. For an isotropic medium with a constant velocity gradient, the relative geometric spreading and parameter sigma, defined in equation 5, coincide: $L_{GS} = \sigma$, and the propagation complexity $c_r$ vanishes. Other cases indicate higher complexity of the wave/ray phenomenon and lead to positive $c_r$ . In cases of multi-arrival stationary solutions, this ray path complexity criterion can be used, for example, as a criterion for accepting/rejecting a given solution. It is reasonable to consider "simpler" paths as more plausible/reliable. Furthermore, in the applications of seismic modeling and imaging, this type of criterion can be used as a weighting factor applied to modeled/imaged seismic events associated with each ray.

## TRAVELTIME VARIATION FOR A PARAXIAL RAY WITH SHIFTED ENDPOINTS

A paraxial ray may be characterized by setting, for example, shifts at the source and receiver locations, $\Delta\mathbf{x}_S$ and $\Delta\mathbf{x}_R$, from the central ray. Given these BC shifts, one can compute the four ray coordinates (RC) $\gamma_i$ as explained in Appendix E, and combine a general paraxial ray $\mathbf{x}_{\text{prx}}(s) = \mathbf{x}(s) + \mathbf{u}(s)$ from the four basic solutions, where the total normal shift is given in equation E6. Then the first and second traveltime variations ($\delta t$ and $\delta^2 t$) between the two endpoints, $S$ and $R$ (provided by equations C1 and C3, respectively), are given by,

$$\Delta t = \delta t + \delta^2 t = \int_S^R \left( L_{\mathbf{x}} \cdot \mathbf{u} + L_{\mathbf{r}} \cdot \dot{\mathbf{u}} \right) ds + \frac{1}{2} \int_S^R \left( \mathbf{u} \cdot L_{\mathbf{xx}} \cdot \mathbf{u} + 2\mathbf{u} \cdot L_{\mathbf{xr}} \cdot \dot{\mathbf{u}} + \dot{\mathbf{u}} \cdot L_{\mathbf{rr}} \cdot \dot{\mathbf{u}} \right) ds \qquad , \qquad (31)$$

where the vectors $L_{\mathbf{x}}$ and $L_{\mathbf{r}}$ are respectively the spatial and directional derivatives of the Lagrangian, $L(s) = dt(s)/ds$, and the second-order tensors $L_{\mathbf{xx}}$, $L_{\mathbf{rr}}$ and $L_{\mathbf{xr}}$, are the spatial, directional and mixed second derivatives of $L(s)$, respectively. At any point along the central ray, vector $\mathbf{u}(\gamma_1, \gamma_2, \gamma_3, \gamma_4, s)$ is a small normal shift between the given paraxial ray $\mathbf{x}_{\text{prx}}(\gamma_1, \gamma_2, \gamma_3, \gamma_4, s)$ and the central ray $\mathbf{x}(s)$. Note that the linear term $\delta t$ in equation 31 prevails. Since the paraxial normal shift $\mathbf{u}(s)$ represents a linear combination of the four basic solutions $\mathbf{u}_i$ of the Jacobi DRT equation, the first traveltime variation $\delta t$ is also a linear combination of the first variations for these four independent solutions, $\delta t_i$. This is not so for the second traveltime variation $\delta^2 t$. Note that there is no need to compute the integrals in equation 31 for each new combination of $\Delta\mathbf{x}_S$ and $\Delta\mathbf{x}_R$ (or each new combination of $\gamma_i$). Rather, one should first compute the first traveltime variations $\delta t_i$ for all basic solutions $\mathbf{u}_i(s)$ (total of four),

$$\delta t_i = L_\mathbf{x} \cdot \mathbf{u}_i + L_\mathbf{r} \cdot \dot{\mathbf{u}}_i \quad , \qquad i = 1, \ldots 4 \qquad , \tag{32}$$

and (optionally) the second traveltime variations $\delta^2 t_{ij}$ for all combinations $\mathbf{u}_i(s)$ and $\mathbf{u}_j(s)$ (total of sixteen),

$$\delta^2 t_{ij} = \frac{1}{2}\left(\mathbf{u}_i \cdot L_{\mathbf{xx}} \cdot \mathbf{u}_j + 2\mathbf{u}_i \cdot L_{\mathbf{xr}} \cdot \dot{\mathbf{u}}_j + \dot{\mathbf{u}}_i \cdot L_{\mathbf{rr}} \cdot \dot{\mathbf{u}}_j\right), \quad i = 1, \ldots 4 \ ; \quad j = 1, \ldots 4 \qquad , \tag{33}$$

where for $i \neq j$ , $\delta^2 t_{ij} \neq \delta^2 t_{ji}$. The weighted terms for $\delta t$ and $\delta^2 t$ are then combined, where the weights are the RC $\gamma_i$ and their products $\gamma_i \gamma_j$ , respectively,

$$\delta t = \sum_{i=1}^{4} \gamma_i \, \delta t_i \qquad , \qquad \delta^2 t = \sum_{i=1}^{4} \sum_{j=1}^{4} \gamma_i \, \gamma_j \, \delta^2 t_{ij} \qquad . \tag{34}$$

Equation 31 provides the incremental paraxial traveltime between the normal cross-section at the source and the normal cross-section at the receiver. The paraxial source and receiver may not belong to these planes. In other words, the paraxial shifts of the source and receiver may have the axial (tangent to the ray) components in addition to the normal ones. To take the axial components into account, we need to add a correction term to the result in equation 31,

$$\Delta t_{\text{srf}} = \delta t_{\text{srf}} + \delta^2 t_{\text{srf}} \approx \delta t_{\text{srf}} = \frac{\delta s_R}{v_{\text{ray},R}} - \frac{\delta s_S}{v_{\text{ray},S}} \quad , \tag{35}$$

where $\delta t_{\text{srf}}$ is the traveltime of the paraxial propagation between the acquisition surfaces and the planes normal to the ray at the source and receiver, and $\delta s_S$ and $\delta s_R$ are tangent paths related to the paraxial source and receiver and provided by equation E1. Note that for this small correction, the quadratic term $\delta^2 t_{\text{srf}}$ in equation 35 is ignored. The total traveltime variation for the paraxial source and receiver is,

$$\Delta t_{\text{tot}} = \delta t + \delta^2 t + \delta t_{\text{srf}} \quad . \tag{36}$$

As mentioned, the first variation $\delta t$ prevails.

## NORMALIZED RELATIVE GEOMETRIC SPREADING AT THE SOURCE

The normalized relative geometric spreading, $L_{GS} / \sigma$, is a suitable unitless dynamic characteristic, used to compute the reliability/plausibility criterion of the resolving stationary path (equation 30). At the start point of the point-source paraxial ray, both the relative geometric spreading and parameter $\sigma$ vanish, but their ratio has a finite limit,

$$\left( \frac{L_{GS}}{\sigma} \right)_S = \frac{1}{v_{\text{phs},S} \sqrt{\lambda_{1,S} \lambda_{2,S}}} \quad , \tag{37}$$

where $\lambda_{1,S}$ and $\lambda_{2,S}$ are the nonzero eigenvalues of the Lagrangian's Hessian at the source, $L_{\mathbf{rr},S}$. The arclength derivative of this value reads,

$$\frac{d}{ds}\left[\frac{L_{GS}}{\sigma}\right]_S = -\frac{v_{J,S}}{4\sqrt{v_{\text{phs},S}\, v_{\text{ray},S}}}\left[2\frac{\dot{v}_{\text{phs},S}}{v_{\text{phs},S}} - \left(\ddot{\mathbf{u}}_{1,S}\times\dot{\mathbf{u}}_{2,S} + \dot{\mathbf{u}}_{1,S}\times\ddot{\mathbf{u}}_{2,S}\right)\cdot\mathbf{r}_S\right] \quad , \quad (38)$$

where $v_{J,S}$ is the conversion velocity defined in equation 28. The units of the slope in equation 38 are the reciprocals of length, $[\mathrm{L}^{-1}]$.

The source-point second derivatives, $\ddot{\mathbf{u}}_{1,S}$ and $\ddot{\mathbf{u}}_{2,S}$, can be obtained from the Jacobi DRT equation (for this we need to open the brackets in equation 12), along with the constraint $\mathbf{u}(s)\cdot\mathbf{r}(s)=0$, and the first and second derivatives of this constraint wrt the arclength of the central ray. The second derivative of the constraint is needed to suppress the singularity of matrix $L_{\mathbf{rr},S}$ when computing $\ddot{\mathbf{u}}_{1,S}$ and $\ddot{\mathbf{u}}_{2,S}$ with the least-squares approach, as we did in Appendix H of Part II. Equations 37 and 38 are useful for testing the numerical results. The detailed derivation is presented in Appendix G.

## CONCLUSIONS

In this part of our study we present an original variational formulation for dynamic ray tracing (DRT), yielding a linear, second-order, Jacobi ODE, to be solved (in Part VII) by the same finite element method that has been applied to the kinematic ray tracing (KRT). The solution of the proposed Jacobi DRT equation for point-source and plane-wave initial conditions (IC) provides a normal shift vector which is a linear combination of four basic vector solutions normal to the ray

direction – defining general paraxial rays. For point-source paraxial rays, only two basic solutions with their corresponding IC are required. The solutions are zero at the source point, normal to each other in a small proximity of the source, but may become collinear at caustic locations. These vector solutions make it possible to compute the ray Jacobian along the stationary ray and therefore, also the (relative) geometric spreading with the ability to identify (and classify) possible caustics. The theory and implementation are valid for 3D smooth heterogeneous general anisotropic media and for all types of wave modes. The challenges of the shear-wave kinematics, related to the branching and algebraic complexity of the shear group velocity surface, are not actual for the shear-wave dynamics, where we assume that both, the central and the paraxial rays belong to the same shear branch.

Using our proposed ray coordinates (RC), we provide the relationship between the corresponding ray Jacobian and the (reciprocal) relative geometric spreading in general anisotropic media. The latter is commonly used as the amplitude factor of the Green's function.

Finally, we propose computing a new dynamic parameter (attribute), the normalized relative geometric spreading, which can be used to compute a qualifying factor/criterion for evaluating the reliability of the resolved stationary path solution. We further provide its value and the arclength derivative at the start point for a point-source paraxial ray, which we found useful for testing the numerical results of Part VII.

## ACKNOWLEDGEMENT

The authors are grateful to Emerson for financial and technical support for this study and for permission to publish its results. The gratitude is extended to Ivan Pšenčík, Einar Iversen,

Michael Slawinski, Alexey Stovas, Vladimir Grechka, and our colleague Beth Orshalimy, whose valuable remarks helped to improve the content and style of this paper.

## APPENDIX A. SPATIAL AND DIRECTIONAL DERIVATIVES OF THE LAGRANGIAN

Using the proposed arclength-related Lagrangian (equation 2 of Part I),

$$L(\mathbf{x},\mathbf{r}) = \frac{dt}{ds} = \frac{\sqrt{\mathbf{r}\cdot\mathbf{r}}}{v_{\text{ray}}(\mathbf{x},\mathbf{r})} \quad , \quad \mathbf{r} \equiv \dot{\mathbf{x}} \equiv \frac{d\mathbf{x}}{ds} \qquad , \qquad \text{(A1)}$$

in this appendix, we derive the first and second derivatives of the proposed Lagrangian $L[\mathbf{x}(s),\mathbf{r}(s)]$ wrt the nodal locations and directions of the ray trajectory, which are the arclength-dependent coefficients of the proposed linear Jacobi DRT equation,

$$\begin{aligned}
L_{\mathbf{x}} &= \frac{\partial L}{\partial \mathbf{x}} = -\frac{\nabla_{\mathbf{x}} v_{\text{ray}} \sqrt{\mathbf{r}\cdot\mathbf{r}}}{v_{\text{ray}}^2} \quad , \quad L_{\mathbf{r}} = \frac{\partial L}{\partial \mathbf{r}} = \frac{\mathbf{r}}{v_{\text{ray}}\sqrt{\mathbf{r}\cdot\mathbf{r}}} - \frac{\nabla_{\mathbf{r}} v_{\text{ray}} \sqrt{\mathbf{r}\cdot\mathbf{r}}}{v_{\text{ray}}^2} \quad , \\
L_{\mathbf{xx}} &= \frac{\partial^2 L}{\partial \mathbf{x}^2} = 2\frac{\nabla_{\mathbf{x}} v_{\text{ray}} \otimes \nabla_{\mathbf{x}} v_{\text{ray}}}{v_{\text{ray}}^3} - \frac{\nabla_{\mathbf{x}}\nabla_{\mathbf{x}} v_{\text{ray}}}{v_{\text{ray}}^2} \quad , \\
L_{\mathbf{xr}} &= L_{\mathbf{rx}}^T = \frac{\partial^2 L}{\partial \mathbf{x}\partial \mathbf{r}} = -\frac{\nabla_{\mathbf{x}} v_{\text{ray}} \otimes \mathbf{r}}{v_{\text{ray}}^2} + 2\frac{\nabla_{\mathbf{x}} v_{\text{ray}} \otimes \nabla_{\mathbf{r}} v_{\text{ray}}}{v_{\text{ray}}^3} - \frac{\nabla_{\mathbf{x}}\nabla_{\mathbf{r}} v_{\text{ray}}}{v_{\text{ray}}^2} \quad , \\
L_{\mathbf{rr}} &= \frac{\mathbf{I} - \mathbf{r}\otimes\mathbf{r}}{v_{\text{ray}}} - \frac{\mathbf{r}\otimes\nabla_{\mathbf{r}} v_{\text{ray}} + \nabla_{\mathbf{r}} v_{\text{ray}} \otimes \mathbf{r}}{v_{\text{ray}}^2} + 2\frac{\nabla_{\mathbf{r}} v_{\text{ray}} \otimes \nabla_{\mathbf{r}} v_{\text{ray}}}{v_{\text{ray}}^3} - \frac{\nabla_{\mathbf{r}}\nabla_{\mathbf{r}} v_{\text{ray}}}{v_{\text{ray}}^2} \quad ,
\end{aligned} \qquad \text{(A2)}$$

where $L_{\mathbf{x}}$, $L_{\mathbf{r}}$ are vectors of length 3, and $L_{\mathbf{xx}}$, $L_{\mathbf{xr}}$, $L_{\mathbf{rx}}$, $L_{\mathbf{rr}}$ are square matrices of dimension 3. Vectors $\nabla_{\mathbf{x}} v_{\text{ray}}$ and $\nabla_{\mathbf{r}} v_{\text{ray}}$ are spatial and directional gradients of the ray velocity, respectively. Tensors $\nabla_{\mathbf{x}}\nabla_{\mathbf{x}} v_{\text{ray}}$ and $\nabla_{\mathbf{r}}\nabla_{\mathbf{r}} v_{\text{ray}}$ are spatial and directional Hessians of the ray velocity and

$\nabla_\mathbf{x}\nabla_\mathbf{r} v_\text{ray}$ and $\nabla_\mathbf{r}\nabla_\mathbf{x} v_\text{ray}$ are the mixed Hessians. Ravve and Koren (2019) provide a computational workflow to establish these gradients and Hessians in smooth heterogeneous general anisotropic media. The first derivatives $L_\mathbf{x}$ and $L_\mathbf{r}$ define the local traveltime gradients, while the second derivatives $L_\mathbf{xx}$, $L_\mathbf{xr}$, $L_\mathbf{rx}$, $L_\mathbf{rr}$ define the local traveltime Hessians, used also in the kinematic analysis.

## APPENDIX B. NECESSARY AND SUFFICIENT CONDITIONS FOR TRAVELTIME MINIMUM

To conclude whether a stationary point is extreme or not, we need to analyse the second traveltime variation, where the traveltime is defined in equation 6. This analysis requires solving the Jacobi DRT equation, whose coefficients are the second derivatives of the Lagrangian $L$ wrt the ray location, $L_\mathbf{xx}$, the ray (group) velocity direction, $L_\mathbf{rr}$, and the mixed derivatives, $L_\mathbf{xr} = L_\mathbf{rx}^T$, obtained in Appendix A.

It is essential to know whether the stationary ray path delivers an extreme traveltime or a saddle point. There are two necessary conditions for a minimum (maximum):

- Matrix $L_\mathbf{rr}$ should be positive semidefinite (negative semidefinite) at any point of the stationary path.
- There should be no caustics along the whole ray path.

The first necessary condition is local, and the second is global. These conditions become also sufficient if positive semi-definiteness (for a minimum) or negative semi-definiteness (for a maximum) are replaced by positive definiteness and negative definiteness, respectively (e.g., Gelfand and Fomin, 2000; Liberzon, 2010).

It follows from the first-order homogeneity property of our proposed Lagrangian, equation 8, that for both isotropic and anisotropic media, zero is one of the eigenvalues of matrix $L_{\mathbf{rr}}$, and the corresponding eigenvector is the ray velocity direction $\mathbf{r}$, i.e., $\det L_{\mathbf{rr}} = 0$, $L_{\mathbf{rr}} \cdot \mathbf{r} = 0$ (Bliss, 1916). Formally, this means that the sufficient conditions for extreme traveltime are never satisfied. We emphasize, however, that the sufficient conditions are not necessary.

For isotropic media, the Lagrangian's directional Hessian matrix reduces to,

$$L_{\mathbf{rr}} = \frac{\mathbf{I} - \mathbf{r} \otimes \mathbf{r}}{v(\mathbf{x})} \quad . \tag{B1}$$

The two nonzero eigenvalues are $1/v(\mathbf{x})$, and the matrix is positive-semidefinite. Note that the eigenvector of the zero eigenvalue is directed along the ray, while the perturbations of the paraxial rays are meaningful only in the plane normal to the central ray. Therefore, due to the special direction of the eigenvector corresponding to the zero eigenvalue, we can treat the positive semidefinite matrix $L_{\mathbf{rr}}$ for isotropic media as positive definite. Thus, for any isotropic media, the local condition for a minimum is always satisfied, making a maximum traveltime path impossible. Obviously, this statement will also hold for weak anisotropy: The two nonzero eigenvalues of matrix $L_{\mathbf{rr}}$ are strictly positive, i.e., for any vector $\boldsymbol{\eta}$,

$$\boldsymbol{\eta} \cdot L_{\mathbf{rr}} \cdot \boldsymbol{\eta} \ge 0 \quad . \tag{B2}$$

The quadratic form vanishes only in the case when vector $\boldsymbol{\eta}$ is collinear with the ray direction $\mathbf{r}$. In anisotropic media, the two different nonzero eigenvalues $\lambda$ of matrix $L_{\mathbf{rr}}$ are the roots of the quadratic equation,

$$\lambda^2 - \operatorname{tr} L_{\mathbf{rr}} \lambda + \frac{1}{2}\left(\operatorname{tr}^2 L_{\mathbf{rr}} - \operatorname{tr} L_{\mathbf{rr}}^2\right) = 0 \qquad , \tag{B3}$$

where $\operatorname{tr}^2 L_{\mathbf{rr}}$ is the squared trace of the matrix, and $\operatorname{tr} L_{\mathbf{rr}}^2$ is the trace of the squared matrix. To obtain the two eigenvalues positive, both the trace $\operatorname{tr} L_{\mathbf{rr}}$ and the difference $\operatorname{tr}^2 L_{\mathbf{rr}} - \operatorname{tr} L_{\mathbf{rr}}^2$ should be positive. These coefficients can be computed with equation A2, where the auxiliary identities given in equation A12 of Part I essentially simplify the final formulae,

$$\frac{\operatorname{tr} L_{\mathbf{rr}}}{2} = \frac{1}{v_{\text{ray}}} + \frac{\nabla_{\mathbf{r}} v_{\text{ray}} \cdot \nabla_{\mathbf{r}} v_{\text{ray}}}{v_{\text{ray}}^3} - \frac{\operatorname{tr} \nabla_{\mathbf{r}} \nabla_{\mathbf{r}} v_{\text{ray}}}{2 v_{\text{ray}}^2} \qquad , \tag{B4}$$

$$\begin{aligned} \frac{\operatorname{tr}^2 L_{\mathbf{rr}} - \operatorname{tr} L_{\mathbf{rr}}^2}{2} = {} & \frac{v_{\text{ray}}^2 + 3 \nabla_{\mathbf{r}} v_{\text{ray}} \cdot \nabla_{\mathbf{r}} v_{\text{ray}}}{v_{\text{ray}}^4} + \frac{\operatorname{tr}^2 \nabla_{\mathbf{r}} \nabla_{\mathbf{r}} v_{\text{ray}} - \operatorname{tr}\left(\nabla_{\mathbf{r}} \nabla_{\mathbf{r}} v_{\text{ray}}\right)^2}{2 v_{\text{ray}}^4} \\ & + 2 \frac{\nabla_{\mathbf{r}} v_{\text{ray}} \cdot \nabla_{\mathbf{r}} \nabla_{\mathbf{r}} v_{\text{ray}} \cdot \nabla_{\mathbf{r}} v_{\text{ray}}}{v_{\text{ray}}^5} - \frac{\operatorname{tr} \nabla_{\mathbf{r}} \nabla_{\mathbf{r}} v_{\text{ray}} \left(v_{\text{ray}}^2 + 2 \nabla_{\mathbf{r}} v_{\text{ray}} \cdot \nabla_{\mathbf{r}} v_{\text{ray}}\right)}{v_{\text{ray}}^5} \quad . \end{aligned} \tag{B5}$$

Yet, it is not easy to draw a conclusion about the sign of the coefficients in equations B4 and B5 without computing them. Although it has not been proven for media with strong anisotropy, one can assume that traveltime maxima are unlikely to exist.

So far we have explored the first (local) necessary condition for the stationary ray to be an extremum (minimum). Next, we check the second (global) condition – the absence of caustics. If one or more caustics exists, the stationary path is a saddle point solution. As already mentioned, traditionally, the conditions for the caustics and their types are derived from the transformation matrix **Q** as a solution of the Hamiltonian-type DRT equation set. In this study, however, the DRT is governed by the Lagrangian-type Jacobi equation. In the next appendix (Appendix C), we follow Bliss (1916) and derive the Jacobi DRT equation for computing the normal shifts (for the given initial conditions) which provide the computation of the geometric spreading and the identification and classification of possible caustics.

Discussion on the minimum criteria

In the discussion below, we summarize the minimum criteria, distinguishing between the known proven facts (paragraph 1) and our assumptions (paragraphs 2 and 3 below).

1) It is known from variational calculus (e.g., Gelfand and Fomin, 2000) that

a) the necessary conditions for a minimum are the positive semidefinite matrix $L_{\mathbf{rr}}$ (at any point along the ray) and the absence of caustics, and

b) the sufficient conditions for a minimum are the positive definite matrix $L_{\mathbf{rr}}$ and the absence of caustics.

The condition for the matrix is local; for the absence of caustics, it is global.

2) Matrix $L_{\mathbf{rr}}$ has a zero eigenvalue corresponding to the third eigenvector which is the ray velocity direction (in both isotropic and anisotropic media); thus, it fails to meet the sufficient criteria. Recall that the first two eigenvectors of $L_{\mathbf{rr}}$ at each point along the ray are vectors in a plane normal to the ray; thus, one can assume that the zero eigenvalue does not "break" the sufficient conditions for a minimum. We therefore suggest simplifying the minimum criteria of the previous paragraph, re-formulating them in the following form: For a minimum traveltime in a general heterogeneous anisotropic model, both necessary and sufficient conditions are the positive semidefinite matrix $L_{\mathbf{rr}}$ with two positive eigenvalues, and the absence of caustics. We note that this (reasonable) assumption has not been explicitly proved.

3) For isotropic media, matrix $L_{\mathbf{rr}}$ is always positive semidefinite: the two nonzero eigenvalues are identical and equal to the reciprocal of the medium velocity, $\lambda_{1,2} = v^{-1}(\mathbf{x})$, $\lambda_3 = 0$. For anisotropic media, the nonzero eigenvalues are different. In either case, the corresponding eigenvectors are in the plane normal to the ray. Due to continuity, matrix $L_{\mathbf{rr}}$ has two positive eigenvalues for weak anisotropy as well. We assume that $L_{\mathbf{rr}}$ has two positive eigenvalues for any strength of anisotropy. In our computational practice, this was always the case. With this assumption, the minimum criteria can be further simplified: For a minimum traveltime in a general heterogeneous anisotropic model, the necessary and sufficient criterion is the absence of caustics along a stationary ray path. The presence of caustics means a saddle point. In other words, we assume that the local necessary and sufficient criterion for a minimum of paragraph 2 above is always satisfied, and only the global criterion remains. This has not been proven either. Note that to be on the safe side, one can always compute the two nonzero eigenvalues of matrix $L_{\mathbf{rr}}$ during the anisotropic DRT and make sure that they are positive.

## APPENDIX C. THE JACOBI DRT EQUATION

The second condition for an extreme (minimum) traveltime path, mentioned in Appendix B – the absence of caustics – is governed by the Jacobi differential equation. This equation (the *accessory* equation) governs the normal distance between the central ray and any paraxial ray vs. the arclength (or any other flow parameter) of the central ray. To derive it, we consider the first and second traveltime variations for a ray traveling between points $S$ and $R$. Recall that the first traveltime variation,

$$\delta t = \int_S^R M_1(s)\, ds, \quad \text{where} \quad M_1 = \delta L(s) = L_{\mathbf{x}} \cdot \delta\mathbf{x} + L_{\mathbf{r}} \cdot \delta\mathbf{r} \quad , \tag{C1}$$

vanishes for a stationary path, leading to the Euler-Lagrange equation set,

$$\frac{d}{ds}\frac{\partial M_1}{\partial \delta\mathbf{r}} = \frac{\partial M_1}{\partial \delta\mathbf{x}} \quad . \tag{C2}$$

The Euler-Lagrange equation governing the first traveltime variation is equation 8 of Part I. The second traveltime variation is then written as,

$$\delta^2 t = \frac{1}{2}\int_S^R M_2(s)\, ds, \quad \text{where} \quad M_2 = \delta^2 L(s) = \delta\mathbf{x} \cdot L_{\mathbf{xx}} \cdot \delta\mathbf{x} + 2\delta\mathbf{x} \cdot L_{\mathbf{xr}} \cdot \delta\mathbf{r} + \delta\mathbf{r} \cdot L_{\mathbf{rr}} \cdot \delta\mathbf{r} \quad . \tag{C3}$$

The analysis of equation C3 makes it possible to conclude whether the stationary path is extreme or not. Note, that the factor 2 for the “mixed” term of the integrand in equation C3 appears due to mutually transposed matrices $L_{\mathbf{xr}}$ and $L_{\mathbf{rx}}$,

$$\delta\mathbf{x} \cdot L_{\mathbf{xr}} \cdot \delta\mathbf{r} + \delta\mathbf{r} \cdot L_{\mathbf{rx}} \cdot \delta\mathbf{x} = 2\delta\mathbf{x} \cdot L_{\mathbf{xr}} \cdot \delta\mathbf{r} = 2\delta\mathbf{r} \cdot L_{\mathbf{rx}} \cdot \delta\mathbf{x} \quad \text{since} \quad L_{\mathbf{rx}} = L_{\mathbf{xr}}^T \tag{C4}$$

The corresponding Jacobi equation set is formulated as,

$$\frac{d}{ds}\frac{\partial M_2}{\partial \delta \mathbf{r}} = \frac{\partial M_2}{\partial \delta \mathbf{x}} \qquad . \tag{C5}$$

It represents the Euler-Lagrange equation applied to the second variation (equation C3) instead of the first variation (equation C1). Combining relationships C3 and C5, and taking into account that $d\delta\mathbf{x}/ds = \delta\mathbf{r}$, we obtain the linear, second-order Jacobi equation set with variable (arclength-dependent) coefficients (Bliss, 1916),

$$\frac{d}{ds}\left( L_{\mathbf{rx}} \cdot \delta\mathbf{x} + L_{\mathbf{rr}} \cdot \frac{d\delta\mathbf{x}}{ds} \right) = L_{\mathbf{xx}} \cdot \delta\mathbf{x} + L_{\mathbf{xr}} \cdot \frac{d\delta\mathbf{x}}{ds} \qquad , \tag{C6}$$

or,

$$\frac{d}{ds}\left( L_{\mathbf{rx}} \cdot \mathbf{u} + L_{\mathbf{rr}} \cdot \dot{\mathbf{u}} \right) = L_{\mathbf{xx}} \cdot \mathbf{u} + L_{\mathbf{xr}} \cdot \dot{\mathbf{u}} \qquad , \tag{C7}$$

where $\mathbf{u} = \delta\mathbf{x}$ stands for the normal shift (normal variation of the path), and $\dot{\mathbf{u}} = d\mathbf{u}/ds = \delta\mathbf{r}$ is the derivative of this normal shift wrt the arclength $s$ of the central ray. We emphasize that $\dot{\mathbf{u}} \cdot \dot{\mathbf{u}} \neq 1$ because $\dot{\mathbf{u}}(s)$ is the derivative of the normal shift $\mathbf{u}(s)$ wrt the arclength $s$ of the central ray $\mathbf{x}(s)$, which is not the arclength $s_{\text{prx}}$ of the paraxial ray $\mathbf{x}_{\text{prx}}(s)$. However, the arclength of the central ray is a convenient flow parameter also for the paraxial ray, and there exists a positive scalar metric that relates the arclengths of the central and paraxial rays,

$$\begin{aligned} \frac{ds_{\text{prx}}}{ds} &= \sqrt{\dot{\mathbf{x}}_{\text{prx}}(s) \cdot \dot{\mathbf{x}}_{\text{prx}}(s)} = \sqrt{\left[\dot{\mathbf{x}}(s) + \dot{\mathbf{u}}(s)\right] \cdot \left[\dot{\mathbf{x}}(s) + \dot{\mathbf{u}}(s)\right]} = \sqrt{\left[\mathbf{r}(s) + \dot{\mathbf{u}}(s)\right] \cdot \left[\mathbf{r}(s) + \dot{\mathbf{u}}(s)\right]} \\ &= \sqrt{\mathbf{r}^2(s) + 2\mathbf{r}(s) \cdot \dot{\mathbf{u}}(s) + \dot{\mathbf{u}}^2(s)} = 1 + \mathbf{r}(s) \cdot \dot{\mathbf{u}}(s) + O\left(\dot{\mathbf{u}}^2\right) = 1 - \dot{\mathbf{r}}(s) \cdot \mathbf{u}(s) + O\left(\dot{\mathbf{u}}^2\right) \qquad , \end{aligned} \tag{C8}$$

where $\dot{\mathbf{r}}(s)$ is the vector-form curvature of the central ray.

Bliss (1916) further notes that a tangent vector solution of the Jacobi set, $\mathbf{u}_t(s) = \rho_t(s)\mathbf{r}(s)$, always exists, where $\rho_t(s)$ is an arbitrary scalar function. This means that the three Cartesian components of the Jacobi vector equation $\mathbf{j}$ (equation C7 with all terms moved to the left-hand side),

$$\mathbf{j} = \frac{d}{ds}\left(L_{\mathbf{rx}} \cdot \mathbf{u} + L_{\mathbf{rr}} \cdot \dot{\mathbf{u}}\right) - \left(L_{\mathbf{xx}} \cdot \mathbf{u} + L_{\mathbf{xr}} \cdot \dot{\mathbf{u}}\right) \quad , \tag{C9}$$

are linearly dependent. Now, assume for a while that vector $\mathbf{u}(s)$ is not necessarily a solution of the Jacobi equation set, such that vector $\mathbf{j}$ does not vanish. Then, it can be shown (Bliss, 1916) that the scalar product $\varsigma_J$ of the vector $\mathbf{j}$ and the ray direction $\mathbf{r}$, $\varsigma_J = \mathbf{j} \cdot \mathbf{r}$, vanishes, and thus, the linear dependence between the components of $\mathbf{j}$ reads, $\varsigma_J = \sum_{i=1}^{3} j_i r_i = 0$. In order to prove this statement, we explicitly compute the scalar product,

$$\begin{aligned}\varsigma_J &= \mathbf{j} \cdot \mathbf{r} = \mathbf{r} \cdot \left[\frac{d}{ds}\left(L_{\mathbf{rx}} \cdot \mathbf{u} + L_{\mathbf{rr}} \cdot \dot{\mathbf{u}}\right) - \left(L_{\mathbf{xx}} \cdot \mathbf{u} + L_{\mathbf{xr}} \cdot \dot{\mathbf{u}}\right)\right] \\ &= \frac{d}{ds}\left[\mathbf{r} \cdot \left(L_{\mathbf{rx}} \cdot \mathbf{u} + L_{\mathbf{rr}} \cdot \dot{\mathbf{u}}\right)\right] - \dot{\mathbf{r}} \cdot \left(L_{\mathbf{rx}} \cdot \mathbf{u} + L_{\mathbf{rr}} \cdot \dot{\mathbf{u}}\right) - \mathbf{r} \cdot \left(L_{\mathbf{xx}} \cdot \mathbf{u} + L_{\mathbf{xr}} \cdot \dot{\mathbf{u}}\right) \quad .\end{aligned} \tag{C10}$$

Using equation 11 we obtain,

$$\begin{aligned}\varsigma_J &= \frac{d}{ds}\left(\mathbf{r} \cdot L_{\mathbf{rx}} \cdot \mathbf{u}\right) - \dot{\mathbf{r}} \cdot \left(L_{\mathbf{rx}} \cdot \mathbf{u} + L_{\mathbf{rr}} \cdot \dot{\mathbf{u}}\right) - \mathbf{r} \cdot \left(L_{\mathbf{xx}} \cdot \mathbf{u} + L_{\mathbf{xr}} \cdot \dot{\mathbf{u}}\right) \\ &= \frac{d}{ds}\left(\mathbf{u} \cdot L_{\mathbf{xr}} \cdot \mathbf{r}\right) - \mathbf{u} \cdot L_{\mathbf{xr}} \cdot \dot{\mathbf{r}} - \dot{\mathbf{u}} \cdot L_{\mathbf{rr}} \cdot \dot{\mathbf{r}} - \mathbf{u} \cdot L_{\mathbf{xx}} \cdot \mathbf{r} - \dot{\mathbf{u}} \cdot L_{\mathbf{rx}}\mathbf{r} \quad .\end{aligned} \tag{C11}$$

Next, we apply equation 10,

$$\begin{aligned}\varsigma_J &= \frac{d}{ds}\left(\mathbf{r}\cdot L_{\mathbf{rx}}\cdot\mathbf{u}\right)-\dot{\mathbf{r}}\cdot\left(L_{\mathbf{rx}}\cdot\mathbf{u}+L_{\mathbf{rr}}\cdot\dot{\mathbf{u}}\right)-\mathbf{r}\cdot\left(L_{\mathbf{xx}}\cdot\mathbf{u}+L_{\mathbf{xr}}\cdot\dot{\mathbf{u}}\right)\\ &=\frac{d}{ds}\left(\mathbf{u}\cdot L_{\mathbf{x}}\right)-\mathbf{u}\cdot L_{\mathbf{xr}}\cdot\dot{\mathbf{r}}-\dot{\mathbf{u}}\cdot L_{\mathbf{rr}}\cdot\dot{\mathbf{r}}-\mathbf{u}\cdot L_{\mathbf{xx}}\cdot\mathbf{r}-\dot{\mathbf{u}}\cdot L_{\mathbf{rx}}\mathbf{r}\quad .\end{aligned} \tag{C12}$$

Note that,

$$\frac{d}{ds}\left(\mathbf{u}\cdot L_{\mathbf{x}}\right)=\dot{\mathbf{u}}\cdot L_{\mathbf{x}}+\mathbf{u}\cdot\frac{dL_{\mathbf{x}}}{ds}=\dot{\mathbf{u}}\cdot L_{\mathbf{x}}+\mathbf{u}\cdot L_{\mathbf{xx}}\cdot\mathbf{r}+\mathbf{u}\cdot L_{\mathbf{xr}}\cdot\dot{\mathbf{r}} \quad . \tag{C13}$$

Introduction of equation C13 into C12 leads to,

$$\varsigma_J=\dot{\mathbf{u}}\cdot L_{\mathbf{x}}-\dot{\mathbf{u}}\cdot L_{\mathbf{rr}}\cdot\dot{\mathbf{r}}-\dot{\mathbf{u}}\cdot L_{\mathbf{rx}}\mathbf{r}=\dot{\mathbf{u}}\cdot L_{\mathbf{x}}-\dot{\mathbf{u}}\cdot\frac{dL_{\mathbf{r}}}{ds}=\dot{\mathbf{u}}\cdot\left(L_{\mathbf{x}}-\frac{dL_{\mathbf{r}}}{ds}\right)=\dot{\mathbf{u}}\cdot\left(L_{\mathbf{x}}-\frac{d\mathbf{p}}{ds}\right) \quad . \tag{C14}$$

Recall that $d\mathbf{p}/ds=L_{\mathbf{x}}=-H_{\mathbf{x}}$ is one of the kinematic ray tracing equations. Thus, the scalar product $\varsigma_J$ vanishes, which means that the three Cartesian components of the Jacobi vector equation set are dependent. Hence, we deduce two consequences that follow from the fact that any tangent solution satisfies the Jacobi DRT equation,

- The tangent solution does not provide additional information. Only the solutions $\mathbf{u}(s)$ which are normal to the ray (normal solutions or normal shifts) are informative. Furthermore, since the shift $\mathbf{u}$ is normal to the central ray at any point along the path, this property can be differentiated wrt the arclength of the central ray, leading to,

$$\mathbf{u}\cdot\mathbf{r}=0 \quad \text{and} \quad \frac{d}{ds}\left(\mathbf{u}\cdot\mathbf{r}\right)=\dot{\mathbf{u}}\cdot\mathbf{r}+\mathbf{u}\cdot\dot{\mathbf{r}}=0 \quad . \tag{C15}$$

- The linear, second-order Jacobi DRT equation set C7 consists of three scalar equations; however, only two of the three equations are independent. Hence, in order to define a paraxial ray, instead of six basic solutions, only four are required, where each of them requires two vector-form initial conditions, normally for $\mathbf{u}_{i,S}$ and $\dot{\mathbf{u}}_{i,S}$ , $i=1, \ldots 4$.

The appearance of caustics (also referred to as conjugate points or singular points) along the ray can be described as intersections of the Jacobi set solution $\mathbf{u}(s)$ with the central ray path. Note that for a point-source paraxial ray, $\mathbf{u}(s)$ consists of a linear combination of only the two basic solutions $\mathbf{u}_1(s)$ and $\mathbf{u}_2(s)$, and a caustic occurs when the two basic solutions become collinear or at least one of them vanishes.

Each single solution $\mathbf{u}(s)$ of the linear, second-order, Jacobi differential equation C7 requires two vector-form initial or boundary conditions. In this study, we consider primarily paraxial rays emerging from a point source, as these rays are needed to compute the geometric spreading and to discover possible caustics (amplitude and phase of the Green function). Consequently, there are no ray path variations at the source, $\mathbf{u}_S=0$. The freedom remains for the second vector-form initial condition associated with $\dot{\mathbf{u}}_S$; thus, the number of independent basic solutions further reduces: there are only two basic point-source solutions, $\mathbf{u}_1$ and $\mathbf{u}_2$. In other words, if we consider all basic solutions, their number is four. If we consider only those basic solutions that vanish at the origin (source), their number is two. The other two basic solutions, $\mathbf{u}_3$ and $\mathbf{u}_4$ for a plane wave, constitute a paraxial ray slightly distant from and parallel to the central ray at the origin. The most general paraxial ray is a combination of the point source and the plane-wave paraxial rays. We note that a combination of point-source and plane-wave initial conditions is only one choice among other options; however, it seems to be the most convenient choice. In either case, the number of basic solutions is four. In the next appendix, we discuss the initial conditions for the basic solutions, considering two different scenarios. In both cases, matrix (tensor) $L_{\mathbf{rr},S}$ has a vanishing eigenvalue, and the two other nonzero (positive) eigenvalues may be a) different or b) identical.

## APPENDIX D. INITIAL CONDITIONS FOR BASIC SOLUTIONS

A general paraxial ray includes a linear combination of the four basic solutions for the DRT equation, $\mathbf{u}_i(s)$, $i=1,\ldots 4$, where each basic solution differs by its initial conditions (IC). We define the IC for each basic solution by setting the values of the DoF of the Jacobi DRT equation at the source point $S$. These DoF are the normal shift vector and its derivative wrt the arclength of the central ray, $\mathbf{u}_{i,S} \equiv \mathbf{u}_i(S)$ and $\dot{\mathbf{u}}_{i,S} \equiv \dot{\mathbf{u}}_i(S)$. Vector $\mathbf{u}_i(s)$ is a component of a general paraxial ray representing an infinitesimal shift in a plane normal to the ray direction at each current point $s$ along the central ray, $\mathbf{u}(s)\cdot\mathbf{r}(s)=0$. Since the Jacobi DRT set is linear, any paraxial ray can be constructed with a linear combination of the four basic solutions.

Different initial conditions (IC) or boundary conditions (BC) (or even mixed conditions) can be set, in order to solve the proposed Jacobi DRT equation and define paraxial rays. Below we present the two commonly used IC: point source and plane wave.

### Initial conditions for a point-source paraxial ray

Distinct eigenvalues of matrix $L_{\mathbf{rr},S}$

To construct a point source paraxial ray, we need two basic solutions, $\mathbf{u}_1$ and $\mathbf{u}_2$, that vanish at the origin. By definition, the shifts $\mathbf{u}_i(s)$ along the ray are normal to the central ray $\mathbf{u}(s)\cdot\mathbf{r}(s)=0$. Since this property holds for any current arclength, the derivative of the left-hand side of this equation, wrt the arclength, also vanishes. The property is formulated in equation 13.

At the point source, the normal vector shifts $\mathbf{u}_{i,S}$ , $i=1,2$, vanish. Thus, it follows from equation 13 that the derivatives of the shifts at the origin are normal to the central ray, $\dot{\mathbf{u}}_{i,S}\cdot\mathbf{r}_S=0$. Since $\mathbf{u}_1$ and $\mathbf{u}_2$ are basic solutions of a general point-source paraxial ray, their derivatives at the source point, $\dot{\mathbf{u}}_{1,S}$ and $\dot{\mathbf{u}}_{2,S}$ are also normal to each other, and the absolute values of their derivatives are normalized to a unit length. The normalized eigenvectors of matrix $L_{\mathbf{rr},S}$, corresponding to the two nonzero eigenvalues, satisfy the abovementioned initial conditions for the derivatives $\dot{\mathbf{u}}_{1,S}$ and $\dot{\mathbf{u}}_{2,S}$. The third eigenvalue of matrix $L_{\mathbf{rr}}$ vanishes (the matrix is noninvertible), and the corresponding eigenvector is $\mathbf{r}$, thus, $L_{\mathbf{rr}}\mathbf{r}=0$. Hence, the initial conditions for the point source read,

$$\begin{aligned} &L_{\mathbf{rr},S}\dot{\mathbf{u}}_{1,S}=\lambda_{1,S}\dot{\mathbf{u}}_{1,S} \ , \quad \dot{\mathbf{u}}_{1,S}\times\dot{\mathbf{u}}_{2,S}\cdot\mathbf{r}_S=1 \ , \\ &L_{\mathbf{rr},S}\dot{\mathbf{u}}_{2,S}=\lambda_{2,S}\dot{\mathbf{u}}_{2,S} \ , \quad \mathbf{u}_{1,S}=\mathbf{u}_{2,S}=0 \quad , \end{aligned} \tag{D1}$$

where $\lambda_{1,S}$ and $\lambda_{2,S}$ are the nonzero (positive) eigenvalues of $L_{\mathbf{rr},S}$. Their corresponding eigenvectors are normal to the ray and to each other. For two different eigenvalues $\lambda_{1,S}$ and $\lambda_{2,S}$ , the corresponding eigenvectors are fully defined. This case is schematically shown in Figure 2a. In some cases, for example, for isotropic media, the two nonzero eigenvalues coincide, $\lambda_{1,S}=\lambda_{2,S}$ , and their eigenvectors are not fully defined (although they are still normal to the ray and to each other). In this special case, referred to as the “double eigenvalue” case, we need to apply a remedy to fully define the eigenvectors normal to the ray. However, before explaining this remedy, we need first to recall some basic definitions of the curved-line geometry, to be later used in the definition of the eigenvectors corresponding to the double eigenvalues. The (ray path) curved line in 3D space can be characterized by its (in-plane measure) curvature and (out-of-plane measure) torsion. At each point $s$ along the ray path, there exists a contact plane of the second-order (i.e., touching, not crossing; unless the path is locally a straight line), to be further referred as the "tangent plane" or the local "ray path plane". The ray path plane includes two vectors: the normalized tangent vector $\mathbf{r}(s)$ and the non-normalized curvature vector $\dot{\mathbf{r}}(s)$ (also called the "normal" vector). The normal to the tangent plane (also called "bi-normal" vector) is the cross-product of these two vectors, $\mathbf{r}(s)\times\dot{\mathbf{r}}(s)$. Both normal and bi-normal vectors are perpendicular to the ray path (to the ray direction $\mathbf{r}$), where the normal vector belongs to the plane of the ray trajectory, and the bi-normal vector is perpendicular to that plane. In other words, the normal and bi-normal vectors represent the in-plane and out-of-plane directions, respectively, both perpendicular to the ray. Thus, these two vectors are suitable for defining the initial conditions of the Jacobi DRT equation, for the normal shift vectors or their derivatives. (We assume here that the curvature does not vanish at the source point. The case with a locally straight ray trajectory in the proximity of the source is considered separately.) A ray tangent, a normal and a bi-normal are schematically shown in Figure 2b.

<u>Double eigenvalue and a curved central ray at the start point</u>

Generally, multiple eigenvalues are characterized by algebraic and geometric multiplicity. The algebraic multiplicity of an eigenvalue is its multiplicity as the root of the characteristic polynomial. The geometric multiplicity is the number of linearly independent eigenvectors corresponding to the multiple eigenvalue (e.g., Nering, 1970; Anton, 1987; Golub and Van Loan, 1996; Burden and Faires, 2005). Geometric multiplicity does not exceed algebraic; it can be only equal or less. For the double nonzero eigenvalues of matrix $L_{\mathbf{rr},S}$, the algebraic and geometric multiplicities are identical: Two linearly independent eigenvectors correspond to this double eigenvalue, both normal to the ray and to each other. However, they are not fully defined (the two mutually orthogonal vectors can rotate about the ray direction $\mathbf{r}$). In the following paragraphs we suggest a way to completely define these vectors.

Note that the ray direction vector $\mathbf{r}(s)$ has a unit length along any point of the path. Thus, it can only change its direction, which means that the curvature vector $\dot{\mathbf{r}}(s)$ is normal to $\mathbf{r}(s)$,

$$\dot{\mathbf{r}}(s) \cdot \mathbf{r}(s) = 0 \quad . \tag{D2}$$

This means that $\dot{\mathbf{u}}_{1,S}$ can be chosen collinear with the ray normal $\dot{\mathbf{r}}_S$, while $\dot{\mathbf{u}}_{2,S}$ is collinear with the bi-normal $\mathbf{r}_S \times \dot{\mathbf{r}}_S$, and the IC of the two basic point-source solutions are,

$$\dot{\mathbf{u}}_{1,S} = \frac{\dot{\mathbf{r}}_S}{|\dot{\mathbf{r}}_S|} \quad , \quad \dot{\mathbf{u}}_2 = \mathbf{r}_S \times \dot{\mathbf{u}}_{1,S} = \frac{\mathbf{r}_S \times \dot{\mathbf{r}}_S}{|\dot{\mathbf{r}}_S|} \quad , \quad \mathbf{u}_{1,S} = \mathbf{u}_{2,S} = 0 \quad . \tag{D3}$$

A case of a double positive eigenvalue and a locally curved path is schematically shown in Figure 2b, for the point-source paraxial ray. Thus, for a point source, the two vector shift derivatives at the origin $\dot{\mathbf{u}}_{1,S}$ and $\dot{\mathbf{u}}_{2,S}$ are normal to the ray direction $\mathbf{r}_S$ at that point. At all other points along the ray path, the vector shift derivatives $\dot{\mathbf{u}}_1(s)$ and $\dot{\mathbf{u}}_2(s)$ are not necessarily normal to the ray; only the vector shifts $\mathbf{u}_i(s)$ are normal to the ray direction $\mathbf{r}$, i.e. $\mathbf{u}_i(s) \cdot \mathbf{r}(s) = 0$ , $i = 1, 2$. Furthermore, only in the infinitesimal proximity of the source, $\mathbf{u}_1(s)$ and $\mathbf{u}_2(s)$ are normal to each other. For all other values of the arclength, the two solutions are normal to the ray but not necessarily to each other. At the extreme cases of caustics, $\mathbf{u}_1(s)$ and $\mathbf{u}_2(s)$ may become collinear; and in this case their cross-product vanishes. A general point source paraxial ray $\mathbf{x}_{\text{prx}}(\gamma_1, \gamma_2, s)$ is defined in equation 19.

**Initial conditions for a plane-wave paraxial ray**

<u>Distinct eigenvalues of matrix $L_{\mathbf{rr},S}$</u>

The plane-wave RC $\gamma_3$ and $\gamma_4$ are defined as the coefficients of the basic normal shifts $\mathbf{u}_3$ and $\mathbf{u}_4$ constituting the fundamental plane-wave solution. At the origin, the paraxial ray $\mathbf{x}_{\text{prx}}$ is collinear with the central ray $\mathbf{x}$. The basic solutions at the origin $\mathbf{u}_{3,S}$ and $\mathbf{u}_{4,S}$ are normal to the central ray and to each other and are normalized to the unit length. Thus, it is reasonable to assign the normalized eigenvectors of matrix $L_{\mathbf{rr},S}$ corresponding to the distinct nonzero eigenvalues as the start-point normal shifts $\mathbf{u}_{3,S}$ and $\mathbf{u}_{4,S}$,

$$L_{\mathbf{rr},S}\mathbf{u}_{3,S} = \lambda_{1,S}\mathbf{u}_{3,S} \quad , \quad L_{\mathbf{rr},S}\mathbf{u}_{4,S} = \lambda_{2,S}\mathbf{u}_{4,S} \quad , \quad \mathbf{u}_{3,S} \times \mathbf{u}_{4,S} \cdot \mathbf{r}_S = 1 \qquad . \qquad \text{(D4)}$$

For a plane wave, the start-point arclength derivatives $\dot{\mathbf{u}}_{3,S}$ and $\dot{\mathbf{u}}_{4,S}$ do not vanish; they are dependent and have to be computed. Since for a plane wave, the central ray $\mathbf{x}$ and the paraxial ray $\mathbf{x}_{\text{prx}}$ are collinear in the proximity of the start point *S*, the two ray directions coincide, $\mathbf{r}_{\text{prx},S} = \mathbf{r}_S$. We apply equation 16 for the paraxial ray, ignoring the high-order (nonlinear) terms.

It holds for any linear combination of $\mathbf{u}_{3,S}$ and $\mathbf{u}_{4,S}$, and thus, it holds for $\mathbf{u}_{3,S}$ and $\mathbf{u}_{4,S}$ separately, which leads to,

$$\dot{\mathbf{u}}_{3,S} = -\left(\dot{\mathbf{r}}_S \cdot \mathbf{u}_{3,S}\right)\mathbf{r}_S \quad , \qquad \dot{\mathbf{u}}_{4,S} = -\left(\dot{\mathbf{r}}_S \cdot \mathbf{u}_{4,S}\right)\mathbf{r}_S \tag{D5}$$

Equations D4 and D6 constitute the initial conditions for a plane wave in the case of distinct eigenvalues. This case is schematically shown in Figure 3a. The derivatives of the plane-wave initial conditions $\dot{\mathbf{u}}_{3,S}$ and $\dot{\mathbf{u}}_{4,S}$ are collinear to the central ray at the start point, $\mathbf{r}_S$. Note that the normalized eigenvector is defined only up to the sign, so we may set the signs of the scalar products in the brackets of equation D6 arbitrarily: These derivatives may point to the central ray direction $\mathbf{r}_S$, or to the opposite direction, $-\mathbf{r}_S$.

<u>Double nonzero eigenvalues and a curved central ray at the start point</u>

In this case, one of the basic solutions at the source, for the plane wave, $\mathbf{u}_{3,S}$, is a vector in the plane of the ray path and normal to the ray. Then, following equation D2, it is collinear with the curvature vector, $\dot{\mathbf{r}}_S$, normal to the path, and the other basic solution, $\mathbf{u}_{4,S}$, is collinear with the bi-normal, $\mathbf{r}_S \times \dot{\mathbf{r}}_S$

$$\mathbf{u}_{3,S} = \frac{\dot{\mathbf{r}}_S}{\left|\dot{\mathbf{r}}_S\right|} \quad , \qquad \mathbf{u}_{4,S} = \frac{\mathbf{r}_S \times \dot{\mathbf{r}}_S}{\left|\dot{\mathbf{r}}_S\right|} \qquad . \tag{D6}$$

Introducing equation D6 into D5, we obtain the arclength derivatives of the plane-wave basic solutions at the start point,

$$\dot{\mathbf{u}}_{3,S} = -\frac{\dot{\mathbf{r}}_S \cdot \dot{\mathbf{r}}_S}{\left|\dot{\mathbf{r}}_S\right|}\mathbf{r}_S \quad , \qquad \dot{\mathbf{u}}_{4,S} = -\frac{\mathbf{r}_S \times \dot{\mathbf{r}}_S \cdot \dot{\mathbf{r}}_S}{\left|\dot{\mathbf{r}}_S\right|}\mathbf{r}_S \qquad , \tag{D7}$$

which simplify to,

$$\dot{\mathbf{u}}_{3,S} = -\left|\dot{\mathbf{r}}_S\right| \mathbf{r}_S \quad , \qquad \dot{\mathbf{u}}_{4,S} = 0 \qquad . \tag{D8}$$

Equations D6 and D8 constitute the initial conditions for a plane wave in the case of the double nonzero (positive) eigenvalues. A case of a double positive eigenvalue and a locally curved path is schematically shown in Figure 3b, for the plane-wave paraxial ray.

A plane-wave paraxial ray $\mathbf{x}_{\text{prx}}$ defined by equation 21, with the IC of equations D4 and D5 (or D6 and D8), will have (in the proximity of the source) a ray velocity, parallel to the ray velocity of the central ray.

**Point-source and plane-wave paraxial rays**

<u>Double eigenvalue and locally straight central ray path</u>

The case of two identical nonzero eigenvalues of the matrix $L_{\mathbf{rr},S}$ and a locally straight ray in the proximity of the start point, where the curvature vector $\dot{\mathbf{r}}_S = 0$, is a special "degenerative" case, typical in homogeneous isotropic media, (e.g., a source in marine environment). In this case the absolute value of the curvature $\left|\dot{\mathbf{r}}_S\right|$ that appears in the denominators of equations D3 and D6, cannot be applied. To overcome this degenerative case, we define an auxiliary "ghost" vector curvature $\dot{\mathbf{r}}_{\tilde{S}}$. Any vector in the plane normal to the ray direction $\mathbf{r}_S$ can be used as a "ghost" curvature vector $\dot{\mathbf{r}}_{\tilde{S}}$, and the algorithm suggested below is one choice out of many. To find this vector, we represent both normalized vectors $\mathbf{r}_S$ and $\dot{\mathbf{r}}_{\tilde{S}}$ in a spherical frame,

$$\mathbf{r}_S = \left[ \sin\theta_S \cos\psi_S \quad \sin\theta_S \sin\psi_S \quad \cos\theta_S \right]$$
$$\dot{\mathbf{r}}_{\tilde{S}} = \left[ \sin\theta_{\tilde{S}} \cos\psi_{\tilde{S}} \quad \sin\theta_{\tilde{S}} \sin\psi_{\tilde{S}} \quad \cos\theta_{\tilde{S}} \right] \quad , \qquad \text{(D9)}$$

where $\theta_S, \psi_S$ are zenith and azimuth angles of the ray direction, while $\theta_{\tilde{S}}, \psi_{\tilde{S}}$ are those of the ghost curvature. The scalar product of the two vectors vanishes, $\mathbf{r}_S \cdot \dot{\mathbf{r}}_{\tilde{S}} = 0$, and this leads to a constraint between the polar angles,

$$\tan\theta_S \tan\theta_{\tilde{S}} \cos\left(\psi_S - \psi_{\tilde{S}}\right) = -1 \quad . \qquad \text{(D10)}$$

The curvature azimuth $\psi_{\tilde{S}}$ can accept any value and may be used as a free parameter; then the curvature zenith angle becomes dependent. The simplest solution is to assume that the two azimuths coincide, $\psi_{\tilde{S}} = \psi_S$, which leads to,

$$\tan\theta_{\tilde{S}} = -\cot\theta_S \quad \rightarrow \quad \theta_{\tilde{S}} = \theta_S \pm \pi/2 \ , \quad 0 \le \theta_{\tilde{S}} < \pi \quad , \qquad \text{(D11)}$$

where the correct sign should be chosen to fit the range of $\theta_{\tilde{S}}$: If $\theta_S < \pi/2$, we apply plus, otherwise minus. We introduce the result into the second equation of set D9 and obtain the Cartesian projections of the ghost curvature $\dot{\mathbf{r}}_{\tilde{S}}$. Taking into account that this vector is normalized, $\left|\dot{\mathbf{r}}_{\tilde{S}}\right| = 1$, we can now use equations D3 and D6 for the IC of paraxial rays for the point source and plane wave, respectively, also in locally homogeneous and isotropic (near the start point) media. Note also that for the (locally) vanishing curvature of the central ray, equation 16 for the paraxial ray direction simplifies to,

$$\mathbf{r}_{\text{prx}}(s) = \mathbf{r}(s) + \dot{\mathbf{u}}(s) + O\left(\dot{\mathbf{u}}^2\right) \quad , \qquad \text{(D12)}$$

which means that for the plane wave with a straight central ray in the proximity of the start point, $\dot{\mathbf{u}}_j = 0$, $j = 3,4$. In this case, the initial conditions for the point-source and plane-wave paraxial rays become,

$$\begin{matrix} \dot{\mathbf{u}}_1 = \dot{\mathbf{r}}_{\tilde{S}} \ , & \dot{\mathbf{u}}_2 = \mathbf{r}_S \times \dot{\mathbf{r}}_{\tilde{S}} \ , & \mathbf{u}_{1,S} = 0 \ , & \mathbf{u}_{2,S} = 0 \ , \\ \mathbf{u}_3 = \dot{\mathbf{r}}_{\tilde{S}} \ , & \dot{\mathbf{u}}_4 = \mathbf{r}_S \times \dot{\mathbf{r}}_{\tilde{S}} \ , & \dot{\mathbf{u}}_{3,S} = 0 \ , & \dot{\mathbf{u}}_{4,S} = 0 \ , \end{matrix} \quad \left|\mathbf{r}_S\right| = \left|\dot{\mathbf{r}}_{\tilde{S}}\right| = 1 \quad . \qquad \text{(D13)}$$

A case of a double positive eigenvalue and a locally straight path is schematically shown in Figures 2c and 3c for the point-source and plane wave paraxial rays, respectively.

Comment: There is an alternative way to define the point-source and plane-wave IC for a locally straight anisotropic ray at the vicinity of the origin It is based on the directional gradient of the ray velocity $\nabla_{\mathbf{r}} v_{\text{ray}}$, which is a vector representing the directional change for the fastest increase of the ray velocity (provided the medium properties are assumed locally homogeneous). This vector is normal to the ray direction $\mathbf{r}$ (see equation A10 of Part I) and is coplanar with the latter and the slowness vector $\mathbf{p}$. Thus, for a locally straight anisotropic ray near the start point, one can set the ghost curvature to be, $\dot{\mathbf{r}}_{\tilde{S}} = \nabla_{\mathbf{r}} v_{\text{ray}}$. Still, for a locally straight isotropic ray, we apply equation set D13.

## APPENDIX E. RAY COORDINATES VS. PARAXIAL SOURCE AND RECEIVER LOCATIONS

In this appendix, we compute the four ray coordinates $\gamma_i$ of a paraxial ray, given the shifts $\Delta\mathbf{x}_R$ and $\Delta\mathbf{x}_S$ between the locations of the paraxial source and receiver and those of the central ray. First, we decompose each of the shifts $\Delta\mathbf{x}_S$ and $\Delta\mathbf{x}_R$ into two components: tangent and normal to the central ray at the source and receiver, respectively. The two tangent components read,

$$\mathbf{r}_S\,\delta s_S \text{ and } \mathbf{r}_R\delta s_R \;,\quad \text{where}\quad \delta s_S = \Delta\mathbf{x}_S \cdot \mathbf{r}_S \text{ and } \delta s_R = \Delta\mathbf{x}_R \cdot \mathbf{r}_R \qquad , \tag{E1}$$

where the tangent paths $\delta s_S$ and $\delta s_R$ may be positive or negative. The other components are the normal shifts $\mathbf{u}_S$ and $\mathbf{u}_R$ between the paraxial and central rays at the source and receiver,

$$\mathbf{u}_S = \mathbf{r}_S \times \Delta\mathbf{x}_S \times \mathbf{r}_S \text{ and } \mathbf{u}_R = \mathbf{r}_R \times \Delta\mathbf{x}_R \times \mathbf{r}_R \qquad . \tag{E2}$$

The general normal shift is a continuous function that includes four components,

$$\sum_{i=1}^{4}\gamma_i \mathbf{u}_{i,S} = \mathbf{u}_S \quad , \qquad \sum_{i=1}^{4}\gamma_i \mathbf{u}_{i,R} = \mathbf{u}_R \qquad . \tag{E3}$$

The basic shifts $\mathbf{u}_{i,S}$ and $\mathbf{u}_{i,R}$ are known values at the source and at the receiver. The resulting normal shifts $\mathbf{u}_S$ and $\mathbf{u}_R$ are also known. The unknown values are only the ray coordinates (RC) $\gamma_i$ . Next, we obtain the four scalar equations from the two vector equations of set E3. For this, we: a) compute the scalar product of each equation with the normal shift vector at the corresponding endpoint, and b) compute the mixed product of each equation with the normal shift vector and ray direction at the corresponding endpoint, as follows,

$$\begin{aligned} &\sum_{i=1}^{4}\gamma_i\,\mathbf{u}_{i,S}\cdot\mathbf{u}_S = \mathbf{u}_S\cdot\mathbf{u}_S \quad , \qquad \sum_{i=1}^{4}\gamma_i\,\mathbf{u}_{i,R}\cdot\mathbf{u}_R = \mathbf{u}_R\cdot\mathbf{u}_R \,, \\ &\sum_{i=1}^{4}\gamma_i\,\mathbf{u}_{i,S}\times\mathbf{u}_S\cdot\mathbf{r}_{\mathrm{S}} = 0 \quad , \qquad \sum_{i=1}^{4}\gamma_i\,\mathbf{u}_{i,R}\times\mathbf{u}_R\cdot\mathbf{r}_{\mathrm{R}} = 0 \quad . \end{aligned} \tag{E4}$$

The coefficients and right-hand sides of the linear set E4 are scalar values. The set can be arranged as,

$$\begin{bmatrix} \mathbf{u}_{1,S}\cdot\mathbf{u}_S & \mathbf{u}_{2,S}\cdot\mathbf{u}_S & \mathbf{u}_{3,S}\cdot\mathbf{u}_S & \mathbf{u}_{4,S}\cdot\mathbf{u}_S \\ \mathbf{u}_{1,S}\times\mathbf{u}_S\cdot\mathbf{r}_S & \mathbf{u}_{2,S}\times\mathbf{u}_S\cdot\mathbf{r}_S & \mathbf{u}_{3,S}\times\mathbf{u}_S\cdot\mathbf{r}_S & \mathbf{u}_{4,S}\times\mathbf{u}_S\cdot\mathbf{r}_S \\ \mathbf{u}_{1,R}\cdot\mathbf{u}_R & \mathbf{u}_{2,R}\cdot\mathbf{u}_R & \mathbf{u}_{3,R}\cdot\mathbf{u}_R & \mathbf{u}_{4,R}\cdot\mathbf{u}_R \\ \mathbf{u}_{1,R}\times\mathbf{u}_R\cdot\mathbf{r}_R & \mathbf{u}_{2,R}\times\mathbf{u}_R\cdot\mathbf{r}_R & \mathbf{u}_{3,R}\times\mathbf{u}_R\cdot\mathbf{r}_R & \mathbf{u}_{3,R}\times\mathbf{u}_R\cdot\mathbf{r}_R \end{bmatrix} \cdot \begin{bmatrix} \gamma_1 \\ \gamma_2 \\ \gamma_3 \\ \gamma_4 \end{bmatrix} = \begin{bmatrix} \mathbf{u}_S\cdot\mathbf{u}_S \\ 0 \\ \mathbf{u}_R\cdot\mathbf{u}_R \\ 0 \end{bmatrix} \quad . \qquad \text{(E5)}$$

Solving set E5, we obtain the RC $\gamma_i$ that relate the paraxial variations of the source/receiver locations $\Delta\mathbf{x}_S$ and $\Delta\mathbf{x}_R$ to the linear combination of the basic solutions $\mathbf{u}_i$ of the Jacobi DRT equation,

$$\mathbf{u}(s) = \sum_{i=1}^{4} \gamma_i \, \mathbf{u}_i(s) \quad . \qquad \text{(E6)}$$

Note that $\mathbf{u}_{1,S} = \mathbf{u}_{2,S} = 0$ (the point-source IC). Thus, equation set E5 simplifies and becomes decoupled. The first two equations include only the plane-wave RC, $\gamma_3$ and $\gamma_4$. We find them, and we solve the two other equations for the point-source RC, $\gamma_1$ and $\gamma_2$.

A particular important case is a point-source paraxial ray. In this case, $\mathbf{u}_S = 0$, and the first two equations of set E5 lead to, $\gamma_3 = \gamma_4 = 0$. The two other equations simplify to,

$$\begin{bmatrix} \mathbf{u}_{1,R}\cdot\mathbf{u}_R & \mathbf{u}_{2,R}\cdot\mathbf{u}_R \\ \mathbf{u}_{1,R}\times\mathbf{u}_R\cdot\mathbf{r}_R & \mathbf{u}_{2,R}\times\mathbf{u}_R\cdot\mathbf{r}_R \end{bmatrix} \cdot \begin{bmatrix} \gamma_1 \\ \gamma_2 \end{bmatrix} = \begin{bmatrix} \mathbf{u}_R\cdot\mathbf{u}_R \\ 0 \end{bmatrix} \quad , \qquad \text{(E7)}$$

that yields,

$$\gamma_1 = +\frac{(\mathbf{u}_R \cdot \mathbf{u}_R)(\mathbf{u}_{2,R} \times \mathbf{u}_R \cdot \mathbf{r}_R)}{(\mathbf{u}_{1,R} \cdot \mathbf{u}_R)(\mathbf{u}_{2,R} \times \mathbf{u}_R \cdot \mathbf{r}_R) - (\mathbf{u}_{2,R} \cdot \mathbf{u}_R)(\mathbf{u}_{1,R} \times \mathbf{u}_R \cdot \mathbf{r}_R)} \quad ,$$
$$\gamma_2 = -\frac{(\mathbf{u}_R \cdot \mathbf{u}_R)(\mathbf{u}_{1,R} \times \mathbf{u}_R \cdot \mathbf{r}_R)}{(\mathbf{u}_{1,R} \cdot \mathbf{u}_R)(\mathbf{u}_{2,R} \times \mathbf{u}_R \cdot \mathbf{r}_R) - (\mathbf{u}_{2,R} \cdot \mathbf{u}_R)(\mathbf{u}_{1,R} \times \mathbf{u}_R \cdot \mathbf{r}_R)} \quad . \qquad \text{(E8)}$$

## APPENDIX F. RAY JACOBIAN, CAUSTIC CRITERIA AND GEOMETRIC SPREADING

Ray Jacobian

The ray Jacobian is the cross-section area of the ray tube, normal to the ray. Its value depends on the choice of the RC. Once the RC are chosen, this value becomes a physical characteristic independent of the reference frame used for its computation (e.g., Cartesian, RCC or WOC). To obtain the ray Jacobian and geometric spreading, we need two basic vector solutions for the point source paraxial ray (solutions for the plane wave are not needed). A single solution $\mathbf{u} = \delta\mathbf{x}$ of the Jacobi DRT equation has units of distance. The signed cross-section area of the ray tube is the signed Jacobian $J(s)$. It represents the determinant of the transform matrix $\mathbf{Q} = \partial\mathbf{x} / \partial\boldsymbol{\gamma}$ between the Cartesian coordinates and the RC, whose columns $\partial\mathbf{x}_{\text{prx}} / \partial\gamma_1$, $\partial\mathbf{x}_{\text{prx}} / \partial\gamma_2$ and $\partial\mathbf{x}_{\text{prx}} / \partial s$, can be obtained from equation E3,

$$\frac{\partial\mathbf{x}_{\text{prx}}(\gamma_1,\gamma_2,s)}{\partial\gamma_1} = \mathbf{u}_1(s) \ , \quad \frac{\partial\mathbf{x}_{\text{prx}}(\gamma_1,\gamma_2,s)}{\partial\gamma_2} = \mathbf{u}_2(s) \ ,$$
$$\frac{\partial\mathbf{x}_{\text{prx}}(\gamma_1,\gamma_2,s)}{\partial s} = \dot{\mathbf{x}}(s) + \gamma_1\,\dot{\mathbf{u}}_1(s) + \gamma_2\,\dot{\mathbf{u}}_2(s)\Big|_{\gamma_1=0,\ \gamma_2=0} = \dot{\mathbf{x}}(s) = \mathbf{r}(s) \ . \qquad \text{(F1)}$$

The ray Jacobian can be arranged as the mixed product of the three column vectors of matrix $\mathbf{Q}$, or equivalency, as the mixed product of the two independent basic solutions, $\mathbf{u}_1$ and $\mathbf{u}_2$, and the ray direction $\mathbf{r}$,

$$J(s) = \frac{\partial \mathbf{x}_{\text{prx}}}{\partial \gamma_1} \times \frac{\partial \mathbf{x}_{\text{prx}}}{\partial \gamma_2} \cdot \frac{\partial \mathbf{x}_{\text{prx}}}{\partial s} = \mathbf{u}_1(s) \times \mathbf{u}_2(s) \cdot \mathbf{r}(s) \quad . \tag{F2}$$

The ray Jacobian $J(s)$ has units of area, the components of the vector solutions $\mathbf{u}_1$ and $\mathbf{u}_2$ have units of distance, and the components of the ray direction $\mathbf{r}$ are unitless. Note that the cross product $\mathbf{u}_1 \times \mathbf{u}_2$ is collinear with the normalized ray direction $\mathbf{r}$, so that $J = \pm|\mathbf{u}_1 \times \mathbf{u}_2|$.

<u>Caustic criteria</u>

Next, we derive the caustic criteria, where the ray paths of the point-source paraxial rays defined by the linear combination $\mathbf{x}_{\text{prx}}(\gamma_1, \gamma_2, s) = \mathbf{x}(s) + \gamma_1 \mathbf{u}_1(s) + \gamma_2 \mathbf{u}_2(s)$ can intersect the central ray. The ray coordinates $\gamma_1$ and $\gamma_2$, do not vanish simultaneously, i.e., for the point source paraxial ray, $\gamma_1^2 + \gamma_2^2 \neq 0$. This means that in the case of a caustic, either $\mathbf{u}_1$ or $\mathbf{u}_2$, or both vanish, or $\mathbf{u}_1$ and $\mathbf{u}_2$ become collinear. The ray Jacobian $J$ vanishes in either of these cases. The caustics differ by their order. The case when both basic solutions vanish simultaneously is the second-order (point) caustic, while all other cases are related to the first-order (line) caustic.

If one basic solution vanishes and the other does not, the direction of the line for the first-order caustic coincides with the non-vanishing solution. If the two solutions become collinear, the direction of the (line) caustic coincides with both of them. The direction of the (line) caustic $\mathbf{r}_L$ is normal to the ray and to a vector $\mathbf{w}_\text{o}$, which is the eigenvector corresponding to the zero eigenvalue of the $3 \times 3$ transposed matrix $\mathbf{Q}^T = [\mathbf{u}_1 \quad \mathbf{u}_2 \quad \mathbf{r}]^T$. The eigenvector $\mathbf{w}_o$ is normal to the ray, and hence,

$$\mathbf{r}_L = \mathbf{w}_o \times \mathbf{r} \quad , \quad \mathbf{w}_o \cdot \mathbf{r} = 0 \quad . \tag{F3}$$

The criterion for the second-order (point) caustic is defined by,

$$c_P = \mathbf{u}_1 \cdot \mathbf{u}_1 + \mathbf{u}_2 \cdot \mathbf{u}_2 = u_1^2 + u_2^2 = 0 \quad , \tag{F4}$$

where $u_1$ and $u_2$ are the absolute values of vectors $\mathbf{u}_1$ and $\mathbf{u}_2$, which means that each normal solution vanishes separately.

In summary, we obtain two independent normal solutions $\mathbf{u}_1$ and $\mathbf{u}_2$ and their derivatives $\dot{\mathbf{u}}_1$ and $\dot{\mathbf{u}}_2$ at any given point along the ray. (With the finite-element implementation discussed in Part VII, the solutions are obtained at the nodes of the finite elements, and the Hermite interpolation is then used for the internal points.) These values make it possible to compute the desired dynamic properties along the stationary ray. The following equations are applied:

- Equation F2 to compute the Jacobian vs. the arclength
- Equation F3 to define the line direction for the first-order (line) caustic
- Equation F4 to detect the second-order (point) caustic

<u>Geometric spreading</u>

Unlike the ray Jacobian, which is not reciprocal (should we swap the source and receiver, the Jacobian changes), the (relative) geometric spreading is reciprocal. It is a more objective dynamic ray characteristic than the ray Jacobian, because the geometric spreading is also independent of the RC choice.

For isotropic media, the relationship between the ray Jacobian and geometric spreading reads (for the choice of the RC used in this study),

$$L_{GS}(s) = v_S \sqrt{\left|J(s)\right|} \qquad , \qquad \text{(F5)}$$

where $v_S$ is the medium velocity at the source. For a general anisotropic case (including isotropic media as a particular case), we first follow Červený (2000) (equation 4.14.46),

$$L_{GS}(s) = \sqrt{\frac{v_{\text{ray}}(s)}{v_{\text{phs}}(s)} \left| \frac{J(s)}{\det \mathbf{P}^{\text{woc}}_{2\times2,S}} \right|} \qquad , \qquad \text{(F6)}$$

where $v_{\text{phs}}(s)$ and $v_{\text{ray}}(s)$ are the phase and ray velocities, respectively, at any current point along the arclength $s$ of the ray path. In particular, this may be the endpoint of the path – the receiver. We recall that the lowercase symbol $s$ is the arclength, while the uppercase symbol $S$ means the source point. The $2\times2$ matrix $\mathbf{P}^{\text{woc}}_{2\times2,S}$ includes the derivatives of the slowness vector of a paraxial ray wrt the RC, computed in the wavefront-orthonormal coordinates (WOC) at the source point $S$. (We will later show that with our choice of RC, the determinant in F6 can be directly computed in the global Cartesian frame.) To compute this $2\times2$ matrix in the WOC frame, we first compute the corresponding $3\times3$ matrix in Cartesian coordinates, consisting of three column vectors,

$$\mathbf{P} = \begin{bmatrix} \dfrac{\partial \mathbf{p}}{\partial \gamma_1} & \dfrac{\partial \mathbf{p}}{\partial \gamma_2} & \dfrac{\partial \mathbf{p}}{\partial s} \end{bmatrix} \equiv \begin{bmatrix} \mathbf{p}_{\gamma,1} & \mathbf{p}_{\gamma,2} & \mathbf{p}_s \end{bmatrix} \qquad , \qquad \text{(F7)}$$

where the arclength $s$ is considered the third RC, in addition to $\gamma_1$ and $\gamma_2$. To compute these derivatives, we refer to the Hamiltonian DRT equations (Červený, 2000, equation 4.2.4), where the flow variable is the traveltime and the form of the Hamiltonian corresponds to traveltime. Recall that our flow variable is the arclength, and in our notations the Hamiltonian DRT equation set can be written as,

$$\dot{\mathbf{w}} = \mathbf{A}\,\mathbf{w} + \mathbf{B}\mathbf{p}_\gamma \;, \qquad \dot{\mathbf{p}}_\gamma = -\mathbf{C}\,\mathbf{w} - \mathbf{D}\mathbf{p}_\gamma \qquad , \tag{F8}$$

where $\mathbf{w} = \partial \mathbf{x}_{\text{prx}} / \partial \gamma$ is the derivative of the paraxial ray position wrt a definite RC $\gamma$ (whose index is omitted here), referred to also as the Hamiltonian DRT solution or the Hamiltonian (paraxial) shift vector. The Hamiltonian DRT solution $\mathbf{w}(s)$ differs from the Lagrangian (Jacobi) normal shift $\mathbf{u}(s)$. As follows from equation 40, the Hamiltonian solution has a tangent counterpart,

$$\mathbf{w} = \mathbf{u} + \rho_t \mathbf{r} \,, \qquad \text{where} \qquad \dot{\rho}_t = \mathbf{u} \cdot \dot{\mathbf{r}} \qquad . \tag{F9}$$

At the start point $S$ of the point-source ray, both Hamiltonian and Lagrangian shifts vanish,

$$\mathbf{w}_S = \mathbf{u}_S = 0 \quad \rightarrow \quad \rho_{t,S} = 0 \qquad . \tag{F10}$$

Note that according to equation F9, the derivative $\dot{\rho}_{t,S} = 0$ due to $\mathbf{u}_S = 0$. Consequently, not only the Hamiltonian and Lagrangian shifts, but their arclength derivatives as well are identical at the start point of the point-source ray,

$$\dot{\mathbf{w}}_S = \dot{\mathbf{u}}_S + \dot{\rho}_{t,S}\mathbf{r}_S + \rho_{t,S}\dot{\mathbf{r}}_S = \dot{\mathbf{u}}_S + \left(\mathbf{u}_S \cdot \dot{\mathbf{r}}_S\right)\mathbf{r}_S + \rho_{t,S}\dot{\mathbf{r}}_S = \dot{\mathbf{u}}_S \qquad . \tag{F11}$$

Introduction of equations F10 and F11 into F9 leads to,

$$\dot{\mathbf{u}}_S = \mathbf{A}_S\,\mathbf{u}_S + \mathbf{B}_S\,\mathbf{p}_{\gamma,S} \quad , \qquad \dot{\mathbf{p}}_\gamma = -\mathbf{C}_S\,\mathbf{u}_S - \mathbf{D}_S\,\mathbf{p}_{\gamma,S} \qquad , \tag{F12}$$

where subscript $S$ emphasizes that the corresponding object is related to the source point. The matrices $\mathbf{A},\mathbf{B},\mathbf{C},\mathbf{D}$ are the $3\times 3$ spatial, slowness and mixed Hessian matrices of the Hamiltonian,

$$\begin{aligned} \mathbf{A} &= \nabla_{\mathbf{p}}\nabla_{\mathbf{x}} H(\mathbf{x},\mathbf{p}) = H_{\mathbf{px}} \quad , \quad \mathbf{B} = \nabla_{\mathbf{p}}\nabla_{\mathbf{p}} H(\mathbf{x},\mathbf{p}) = H_{\mathbf{pp}} \quad , \\ \mathbf{C} &= \nabla_{\mathbf{x}}\nabla_{\mathbf{x}} H(\mathbf{x},\mathbf{p}) = H_{\mathbf{xx}} \quad , \quad \mathbf{D} = \nabla_{\mathbf{x}}\nabla_{\mathbf{p}} H(\mathbf{x},\mathbf{p}) = H_{\mathbf{xp}} \quad , \end{aligned} \tag{F13}$$

where $H(\mathbf{x},\mathbf{p})$ is the arclength-related Hamiltonian, explained by equations 11 and 13 of Part I. According to the initial conditions, described in Appendix D, all paraxial rays depart from the same origin of the source-point central ray, $\mathbf{u}_S = \partial\mathbf{x}_{\text{prx},S} / \partial\gamma = 0$. We apply a mnemonic notation $\mathbf{B} \equiv H_{\mathbf{pp}}$, and the first equation of set F12 simplifies to,

$$\mathbf{p}_{\gamma,i,S} = H_{\mathbf{pp}}^{-1}(\mathbf{x}_S,\mathbf{p}_S)\,\dot{\mathbf{u}}_{i,S} \quad , \qquad i = 1,2 \qquad , \tag{F14}$$

where we assume that matrix $H_{\mathbf{pp},S} = H_{\mathbf{pp}}(\mathbf{x}_S,\mathbf{p}_S)$ at the source point is invertible. The third column of matrix $\mathbf{P}$ in equation F7 can be obtained from the kinematics,

$$\mathbf{p}_s = \frac{d\mathbf{p}}{ds} = -\frac{\partial H(\mathbf{x},\mathbf{p})}{\partial\mathbf{x}} \equiv -H_{\mathbf{x}}(\mathbf{x},\mathbf{p}) \qquad . \tag{F15}$$

We will see later that this third column is not needed for the computation of matrix $\mathbf{P}_{2\times 2,S}^{\text{woc}}$.

The arclength derivatives of the two basic solutions for the normal paraxial vector shifts $\dot{\mathbf{u}}_{1,S}$ and $\dot{\mathbf{u}}_{2,S}$ are the known normalized vectors (and specified as the initial conditions at the source point, see Appendix D for details).

Next, we rotate matrix $\mathbf{P}$ to the WOC frame. Since we use this frame at the source point only, it can be viewed as just a different Cartesian frame, whose local $x_3$ axis coincides with the slowness direction of the central ray at this point. The rotation is generally defined by the three Euler's angles: azimuth, zenith and spin. In this case, the zenith $\theta_{p,S}$ and azimuth $\psi_{p,S}$ characterize the source slowness direction in the global frame, while the spin can be arbitrary. The most reasonable assumption is to set a zero spin, and the global-to-WOC rotation matrix reads,

$$\mathbf{A}_{\text{rot}} = \begin{bmatrix} +\cos\theta_{p,S}\cos\psi_{p,S} & +\cos\theta\sin\psi_{p,S} & -\sin\theta_{p,S} \\ -\sin\psi_{p,S} & +\cos\psi_{p,S} & 0 \\ +\sin\theta_{p,S}\cos\psi_{p,S} & +\sin\theta_{p,S}\sin\psi_{p,S} & +\cos\theta_{p,S} \end{bmatrix} \quad , \qquad \text{(F16)}$$

where,

$$\cos\theta_{p,S} = \frac{p_{3,S}}{\sqrt{p_{1,S}^2 + p_{2,S}^2 + p_{3,S}^2}} \quad , \qquad \sin\theta_{p,S} = \frac{\sqrt{p_{1,S}^2 + p_{2,S}^2}}{\sqrt{p_{1,S}^2 + p_{2,S}^2 + p_{3,S}^2}} \quad ,$$
$$\cos\psi_{p,S} = \frac{p_{1,S}}{\sqrt{p_{1,S}^2 + p_{2,S}^2}} \quad , \qquad \sin\psi_{p,S} = \frac{p_{2,S}}{\sqrt{p_{1,S}^2 + p_{2,S}^2}} \quad . \qquad \text{(F17)}$$

The third row of the rotation matrix represents the slowness direction at the source point, as viewed from the global frame. Note that matrix $\mathbf{P}$ is not a physical tensor, it is just a set of three columns, which are rotated as separate vectors,

$$\mathbf{P}_{3\times3,S}^{\mathrm{woc}} = \left[\mathbf{A}_{\mathrm{rot}} H_{\mathbf{pp}}^{-1} \dot{\mathbf{u}}_1 \quad \mathbf{A}_{\mathrm{rot}} H_{\mathbf{pp}}^{-1} \dot{\mathbf{u}}_2 \quad -\mathbf{A}_{\mathrm{rot}} H_{\mathbf{x}}\right]_S \quad . \tag{F18}$$

The third row and the third column are then discarded to obtain the $2\times2$ matrix $\mathbf{P}_{2\times2,S}^{\mathrm{woc}}$; thus $H_{\mathbf{x}}\left(\mathbf{x}_S, \mathbf{p}_S\right)$ in equation F18 is not needed. We now compute the determinant of $\mathbf{P}_{2\times2,S}^{\mathrm{woc}}$ in two steps. First, we just discard the third column and obtain the $3\times2$ matrix,

$$\mathbf{P}_{3\times2,S}^{\mathrm{woc}} = \left[\mathbf{A}_{\mathrm{rot}} H_{\mathbf{pp}}^{-1} \dot{\mathbf{u}}_1 \quad \mathbf{A}_{\mathrm{rot}} H_{\mathbf{pp}}^{-1} \dot{\mathbf{u}}_2\right]_S \quad . \tag{F10}$$

Next, we truncate the third row and compute the determinant of the remaining $2\times2$ submatrix. This truncation and computation of the determinant can be arranged as,

$$\det \mathbf{P}_{2\times2,S}^{\mathrm{woc}} = \left(\mathbf{A}_{\mathrm{rot}} H_{\mathbf{pp}}^{-1} \dot{\mathbf{u}}_1\right)_S \times \left(\mathbf{A}_{\mathrm{rot}} H_{\mathbf{pp}}^{-1} \dot{\mathbf{u}}_2\right) \cdot \frac{\mathbf{A}_{\mathrm{rot}} \mathbf{p}_S}{\sqrt{\mathbf{p}_S \cdot \mathbf{p}_S}} \quad . \tag{F20}$$

Indeed, the desired value of $\det \mathbf{P}_{2\times2,S}^{\mathrm{woc}}$ can be obtained as $\det \mathbf{P}_{2\times2,S}^{\mathrm{woc}} = a_1 b_2 - a_2 b_1$ , computed from the $3\times2$ matrix,

$$\begin{bmatrix} a_1 & b_1 \\ a_2 & b_2 \\ a_3 & b_3 \end{bmatrix} \quad , \tag{F21}$$

where $a_i$ and $b_i$ are components of the vectors in the brackets in equation F20. The third factor on the right side of equation F20 has only an $x_3$-component, which is 1 (this is the slowness direction in the local, or WOC frame). In other words,

$$\det \begin{bmatrix} a_1 & b_1 \\ a_2 & b_2 \end{bmatrix} = a_1 b_2 - a_2 b_1 =$$

$$\{a_1 \quad a_2 \quad a_3\} \times \{b_1 \quad b_2 \quad b_3\} \cdot \{0 \quad 0 \quad 1\} = \mathbf{a} \times \mathbf{b} \cdot \frac{\mathbf{p}^{\text{loc}}}{\sqrt{\mathbf{p}^{\text{loc}} \cdot \mathbf{p}^{\text{loc}}}} \quad . \tag{F22}$$

We note that a mixed (triple) product of three vectors represents a volume of a parallelepiped, built on these vectors. This volume is a physical scalar, invariant in any frame, which makes rotation to the WOC unnecessary. Hence, the determinant that we need can be directly computed in the global Cartesian frame. With the introduction of the slowness direction (in the global frame) at the source,

$$\mathbf{n}_S = \frac{\mathbf{p}_S}{\sqrt{\mathbf{p}_S \cdot \mathbf{p}_S}} \quad , \tag{F23}$$

equation F20 simplifies to,

$$\det \mathbf{P}_{2\times2,S}^{\text{woc}} = \left(H_{\mathbf{pp}}^{-1} \dot{\mathbf{u}}_1\right)_S \times \left(H_{\mathbf{pp}}^{-1} \dot{\mathbf{u}}_2\right)_S \cdot \mathbf{n}_S \quad . \tag{F23}$$

Since the WOC frame is not needed to compute the value on the left side of equation F23, we use in the body of the paper a more suitable notation, $v_{J,S} = \dfrac{1}{\sqrt{\left|\det \mathbf{P}_{2\times2,S}^{\text{woc}}\right|}}$. We call parameter $v_{J,S}$ the conversion velocity,

$$v_{J,S} = \frac{1}{\sqrt{\left|\left(H_{\mathbf{pp}}^{-1} \dot{\mathbf{u}}_1\right)_S \times \left(H_{\mathbf{pp}}^{-1} \dot{\mathbf{u}}_2\right)_S \cdot \mathbf{n}_S\right|}} \quad . \tag{F24}$$

It has the units of velocity, where subscript $J$ indicates that it is related to the conversion of the arclength related ray Jacobian into the geometric spreading, and subscript $S$ indicates that its computation is solely performed at the source point. For a particular case of an isotropic medium, equation F24 leads to $v_{J,S} = v_S$, where $v_S$ is just the medium velocity at the source, and equation F6 reduces to equation F5.

Recall that with our choice of the RC, the initial conditions for derivatives of the normal shift vector, $\dot{\mathbf{u}}_{1,S}$ and $\dot{\mathbf{u}}_{2,S}$, are the eigenvectors of matrix $L_{\mathbf{rr},S}$ corresponding to the nonzero eigenvalues, $\lambda_{1,S}$ and $\lambda_{2,S}$. The eigensystems of matrices $L_{\mathbf{rr}}$ and $H_{\mathbf{pp}}^{-1}$ are similar. Both matrices have the same eigenvectors: one of them is the ray direction $\mathbf{r}$, and the two others are in the plane normal to the ray (they are, of course, also normal to each other). As mentioned, the eigenvalues $\lambda_1$ and $\lambda_2$, corresponding to the eigenvectors in the normal plane, are identical for matrices $L_{\mathbf{rr}}$ and $H_{\mathbf{pp}}^{-1}$. The only difference is the third eigenvalue, corresponding to the eigenvector tangent to the ray. Its value is zero for $L_{\mathbf{rr}}$ and nonzero $\lambda_{\mathbf{r}}$ for $H_{\mathbf{pp}}^{-1}$. (Note also that matrices $H_{\mathbf{pp}}$ and $H_{\mathbf{pp}}^{-1}$ have the same eigenvectors, and their eigenvalues are reciprocals of each other.) Taking the above statements into account, we obtain,

$$H_{\mathbf{pp},S}^{-1}\dot{\mathbf{u}}_{1,S} = \lambda_{1,S}\dot{\mathbf{u}}_{1,S} \quad , \quad H_{\mathbf{pp},S}^{-1}\dot{\mathbf{u}}_{2,S} = \lambda_{2,S}\dot{\mathbf{u}}_{2,S} \quad . \qquad \text{(F25)}$$

The eigenvalues $\lambda_1$ and $\lambda_2$ are normally positive (otherwise a saddle point stationary traveltime might be possible even in the absence of caustics; we are unaware of such a case). Introduction of equation F25 into F24 leads to,

$$v_{J,S} = \frac{1}{\sqrt{\lambda_{1,S}\lambda_{2,S} \left| \dot{\mathbf{u}}_{1,S} \times \dot{\mathbf{u}}_{2,S} \cdot \mathbf{n}_S \right|}} \quad . \tag{F26}$$

Both vectors $\dot{\mathbf{u}}_{1,S}$ and $\dot{\mathbf{u}}_{2,S}$ have a unit length, they are normal to the ray and to each other, therefore, their cross-product is

$$\dot{\mathbf{u}}_{1,S} \times \dot{\mathbf{u}}_{2,S} = \pm \mathbf{r}_S \tag{F27}$$

We may assume any sign on the right side of equation F27, and we introduce equation F27 into equation F26. After that the absolute value operator can be removed, and the conversion velocity simplifies to,

$$v_{J,S} = \frac{1}{\sqrt{\lambda_{1,S}\lambda_{2,S}\, \mathbf{r}_S \cdot \mathbf{n}_S}} \quad , \quad \lambda_{1,S} > 0 \quad , \quad \lambda_{2,S} > 0 \quad . \tag{F28}$$

Note that $\mathbf{r}_S \cdot \mathbf{n}_S$ is the (positive) scalar product of the ray and phase directions (both normalized to the unit length), so that,

$$\mathbf{r}_S \cdot \mathbf{n}_S = \cos\beta_S = \frac{v_{\text{phs},S}}{v_{\text{ray},S}} \quad , \tag{F29}$$

where $\beta_S$ is the angle between the phase and ray velocity vectors at the source. The conversion velocity becomes,

$$v_{J,S} = \frac{1}{\sqrt{\lambda_{1,S}\lambda_{2,S}}} \sqrt{\frac{v_{\text{ray},S}}{v_{\text{phs},S}}} \quad . \tag{F30}$$

Combining equation F30 with equation F6 for the geometric spreading, we obtain,

$$L_{GS}\left(s\right)=\sqrt{\frac{v_{\text{ray},S}}{v_{\text{phs},S}}\frac{v_{\text{ray}}\left(s\right)}{v_{\text{phs}}\left(s\right)}\frac{\left|J\left(s\right)\right|}{\lambda_{1,S}\lambda_{2,S}}} \quad . \tag{F31}$$

This relationship is valid for any point along the ray path, in particular, for the receiver.

## APPENDIX G. NORMALIZED GEOMETRIC SPREADING OF A POINT-SOURCE PARAXIAL RAY AT THE START POINT

In this appendix, we compute the normalized geometric spreading of a point-source paraxial ray, $L_{GS}/\sigma$ and its arclength derivative, $d\left(L_{GS}/\sigma\right)/ds$, at the start point.

The normalized geometric spreading at the source point

At (exactly) the source point, both the geometric spreading and parameter $\sigma$ vanish, but their ratio remains finite; this has been confirmed by all our numerical tests. It is interesting to predict the analytical value of the so-called normalized geometric spreading at the source $\left(L_{GS}/\sigma\right)_S$ , and to compare it with our numerical results. For this, we arrange equation F6 as,

$$L_{GS}\left(s\right)=v_{J,S}\sqrt{\frac{v_{\text{ray}}\left(s\right)}{v_{\text{phs}}\left(s\right)}J\left(s\right)} \quad , \tag{G1}$$

where the conversion velocity $v_{J,S}$ is a constant (arclength-independent) factor, and differentiate the logarithm of the two sides of this equation,

$$\frac{\dot{L}_{GS}\left(s\right)}{L_{GS}\left(s\right)}=\frac{1}{2}\frac{\dot{v}_{\text{ray}}\left(s\right)}{v_{\text{ray}}\left(s\right)}-\frac{1}{2}\frac{\dot{v}_{\text{phs}}\left(s\right)}{v_{\text{phs}}\left(s\right)}+\frac{1}{2}\frac{\dot{J}\left(s\right)}{J\left(s\right)} \quad . \tag{G2}$$

Without any loss of generality, we assume here that for infinitesimal values of the arclength, the ray Jacobian is positive (otherwise we just swap the basic solutions $\mathbf{u}_1(s)$ and $\mathbf{u}_2(s)$ in their cross product). Next, we rearrange equation G2,

$$\dot{L}_{GS}(s) = \lim_{ds\to 0}\left\{\frac{v_{J,S}}{2}\sqrt{\frac{v_{\text{ray}}(s)}{v_{\text{phs}}(s)}J(s)}\left[\frac{\dot{v}_{\text{ray}}(s)}{v_{\text{ray}}(s)}-\frac{\dot{v}_{\text{phs}}(s)}{v_{\text{phs}}(s)}\right]+\frac{v_{J,S}}{2}\sqrt{\frac{v_{\text{ray}}(s)}{v_{\text{phs}}(s)}}\,\frac{\dot{J}(s)}{\sqrt{J(s)}}\right\} \quad . \quad \text{(G3)}$$

The point-source ray Jacobian is infinitesimal near the origin, and equation G3 reduces to,

$$\dot{L}_{GS}(s) = \frac{v_{J,S}}{2}\sqrt{\frac{v_{\text{ray},S}}{v_{\text{phs},S}}}\lim_{ds\to 0}\frac{\dot{J}(s)}{\sqrt{J(s)}} \quad , \quad \text{(G4)}$$

where $J = \mathbf{u}_1 \times \mathbf{u}_2 \cdot \mathbf{r}$. Recall that at the origin, $\mathbf{u}_{1,S} = \mathbf{u}_{2,S} = 0$. Thus, in the proximity of the source point,

$$\mathbf{u}_1 = \dot{\mathbf{u}}_1 ds \ , \quad \mathbf{u}_2 = \dot{\mathbf{u}}_2 ds \ , \quad \mathbf{r} = \mathbf{r}_S + \dot{\mathbf{r}}_S ds \quad . \quad \text{(G5)}$$

This leads to,

$$\begin{gathered}\lim_{ds\to 0}\frac{\dot{J}(s)}{\sqrt{J(s)}} = \lim_{ds\to 0}\frac{\dot{\mathbf{u}}_1\times\mathbf{u}_2\cdot\mathbf{r}+\mathbf{u}_1\times\dot{\mathbf{u}}_2\cdot\mathbf{r}+\mathbf{u}_1\times\mathbf{u}_2\cdot\dot{\mathbf{r}}}{\sqrt{\mathbf{u}_1\times\mathbf{u}_2\cdot\mathbf{r}}} = \\ = \frac{2\dot{\mathbf{u}}_1\times\dot{\mathbf{u}}_2\cdot\mathbf{r}\,ds+\dot{\mathbf{u}}_1\times\dot{\mathbf{u}}_2\cdot\dot{\mathbf{r}}\,ds^2}{\sqrt{\dot{\mathbf{u}}_1\times\dot{\mathbf{u}}_2\cdot\mathbf{r}\,ds^2}} = 2\sqrt{\dot{\mathbf{u}}_1\times\dot{\mathbf{u}}_2\cdot\mathbf{r}} = 2 \ .\end{gathered} \quad \text{(G6)}$$

We recall that the derivatives $\dot{\mathbf{u}}_1$ and $\dot{\mathbf{u}}_2$ are normal to the ray direction $\mathbf{r}$ and to each other; in addition, all three vectors have unit lengths, so that their mixed product, $\dot{\mathbf{u}}_1\times\dot{\mathbf{u}}_2\cdot\mathbf{r}=1$. Equation G4 simplifies to,

$$\dot{L}_{GS,S} = v_{J,S} \sqrt{\frac{v_{\mathrm{ray},S}}{v_{\mathrm{phs},S}}} \quad . \qquad \text{(G7)}$$

Likewise, the ray Jacobian and geometric spreading, parameter $\sigma$ also vanishes at the origin, and $\dot{\sigma}_S = v_{\mathrm{ray},S}$. Applying L'Hospital rule, we obtain,

$$\lim_{ds \to 0} \frac{L_{GS}(ds)}{\sigma(ds)} = \left(\frac{L_{GS}}{\sigma}\right)_S = \frac{\dot{L}_{GS}}{\dot{\sigma}_S} = \frac{v_{J,S}}{\sqrt{v_{\mathrm{phs},S}\, v_{\mathrm{ray},S}}} \quad . \qquad \text{(G8)}$$

Thus, the normalized geometric spreading at the source is equal to the ratio between the conversion velocity and the geometric average of the ray and phase velocities at the source. This value is always 1 for isotropic media.

Introduction of equation F30 into G8 leads to an alternative relationship for the normalized geometric spreading at the source,

$$\left(\frac{L_{GS}}{\sigma}\right)_S = \frac{1}{v_{\mathrm{phs},S}\sqrt{\lambda_{1,S}\lambda_{2,S}}} \quad . \qquad \text{(G9)}$$

Slope of the normalized geometric spreading at the source point

A similar analysis (but finer, with higher derivatives) can be applied to compute the slope of the graph, for the normalized geometric spreading in the proximity of the source, $\lim\limits_{s \to S} \dfrac{d}{ds} \dfrac{L_{GS}(s)}{\sigma(s)}$.

Analytical computation of this value may be useful to validate the correctness of the finite-element results (described and implemented in Part VI). It follows from equation G1,

$$\frac{L_{GS}(s)}{\sigma(s)} = \frac{v_{J,S}}{\sigma(s)} \sqrt{\frac{v_{\text{ray}}(s)}{v_{\text{phs}}(s)} J(s)} \quad , \tag{G10}$$

so that,

$$\frac{\frac{d}{ds}\left[L_{GS}(s)/\sigma(s)\right]}{L_{GS}(s)/\sigma(s)} = -\frac{\dot{\sigma}(s)}{\sigma(s)} + \frac{1}{2}\frac{\dot{v}_{\text{ray}}(s)}{v_{\text{ray}}(s)} - \frac{1}{2}\frac{\dot{v}_{\text{phs}}(s)}{v_{\text{phs}}(s)} + \frac{1}{2}\frac{\dot{J}(s)}{J(s)} \quad , \tag{G11}$$

or,

$$\frac{d}{ds}\left[\frac{L_{GS}}{\sigma}\right]_S \frac{\sqrt{v_{\text{phs}}(s) v_{\text{ray}}(s)}}{v_{J,S}} = -\frac{\dot{\sigma}(s)}{\sigma(s)} + \frac{1}{2}\frac{\dot{v}_{\text{ray}}(s)}{v_{\text{ray}}(s)} - \frac{1}{2}\frac{\dot{v}_{\text{phs}}(s)}{v_{\text{phs}}(s)} + \frac{1}{2}\frac{\dot{J}(s)}{J(s)} \quad . \tag{G12}$$

Expand parameter $\sigma$ and the components of the ray Jacobian up to quadratic terms, and the ray and phase velocities up to linear terms,

$$\begin{array}{ll} \mathbf{u}_1(ds) = \dot{\mathbf{u}}_{1,S}\, ds + \ddot{\mathbf{u}}_{1,S}\, \frac{ds^2}{2} + O\left(ds^3\right) & \sigma(ds) = v_{\text{ray},S}\, ds + \dot{v}_{\text{ray},S}\, \frac{ds^2}{2} + O\left(ds^3\right) \\ \mathbf{u}_2(ds) = \dot{\mathbf{u}}_{2,S}\, ds + \ddot{\mathbf{u}}_{2,S}\, \frac{ds^2}{2} + O\left(ds^3\right) & v_{\text{ray}}(s) = v_{\text{ray},S} + \dot{v}_{\text{ray},S}\, ds + O\left(ds^2\right) \\ \mathbf{r}(ds) = \mathbf{r}_S + \dot{\mathbf{r}}_S ds + O\left(ds^2\right) & v_{\text{phs}}(s) = v_{\text{phs},S} + \dot{v}_{\text{phs},S}\, ds + O\left(ds^2\right) \end{array} \quad . \tag{G13}$$

Introducing equation G13 into G12 and applying the limit, we obtain,

$$\begin{gathered} \frac{d}{ds}\left[\frac{L_{GS}}{\sigma}\right]_S \frac{\sqrt{v_{\text{phs}}(s) v_{\text{ray}}(s)}}{v_{J,S}} = \\ -\frac{1}{2}\frac{\dot{v}_{\text{phs}}(s)}{v_{\text{phs}}(s)} + \frac{2\dot{\mathbf{u}}_{1,S} \times \dot{\mathbf{u}}_{2,S} \cdot \dot{\mathbf{r}}_S + \ddot{\mathbf{u}}_{1,S} \times \dot{\mathbf{u}}_{2,S} \cdot \mathbf{r}_S + \dot{\mathbf{u}}_{1,S} \times \ddot{\mathbf{u}}_{2,S} \cdot \mathbf{r}_S}{4\dot{\mathbf{u}}_{1,S} \times \dot{\mathbf{u}}_{2,S} \cdot \mathbf{r}_S} \quad . \end{gathered} \tag{G14}$$

Recall that vector $\mathbf{r}$ is normalized, $\mathbf{r}\cdot\mathbf{r}=1$, and thus, $\dot{\mathbf{r}}\cdot\mathbf{r}=0$. In other words, the curvature vector $\dot{\mathbf{r}}$ is normal to the ray direction, and thus, it is coplanar with the initial conditions $\dot{\mathbf{u}}_{1,S}$ and $\dot{\mathbf{u}}_{2,S}$, which leads to, $\dot{\mathbf{u}}_{1,S}\times\dot{\mathbf{u}}_{2,S}\cdot\dot{\mathbf{r}}_S=0$. Recall also that the mixed product in the denominator is 1, and equation G14 simplifies to,

$$\frac{d}{ds}\left[\frac{L_{GS}}{\sigma}\right]_S=-\frac{v_{J,S}}{4\sqrt{v_{\text{phs},S}\,v_{\text{ray},S}}}\left[2\frac{\dot{v}_{\text{phs},S}}{v_{\text{phs},S}}-\left(\ddot{\mathbf{u}}_{1,S}\times\dot{\mathbf{u}}_{2,S}+\dot{\mathbf{u}}_{1,S}\times\ddot{\mathbf{u}}_{2,S}\right)\cdot\mathbf{r}_S\right] \quad . \qquad \text{(G15)}$$

The units of the tilt in equation G15 are the reciprocals of length, $\left[\mathrm{L}^{-1}\right]$.

The source-point second derivatives $\ddot{\mathbf{u}}_{1,S}$ and $\ddot{\mathbf{u}}_{2,S}$, can be obtained from the Jacobi DRT equation (for this we need to open the brackets in equation 12), along with the constraint $\mathbf{u}(s)\cdot\mathbf{r}(s)=0$, and the first and second derivatives of this constraint wrt the arclength of the central ray.

## LIST OF FIGURES

Figure 1. Scheme of the ray tube for point-source paraxial rays: $\mathbf{r}(s)$ – central ray direction, $\mathbf{u}_1(s)$, $\mathbf{u}_2(s)$ – paraxial shift vectors, normal to $\mathbf{r}(s)$, $\theta_u(s)$ – angle between these vectors, $s$ – arclength of the central ray (red line). At the source point, the normal shifts $\mathbf{u}_1$ and $\mathbf{u}_2$ vanish. In the proximity of the source point, these vectors are infinitesimal and normal to each other. At the line caustic, the normal shift vectors may become collinear. The shaded area of the parallelogram is equal to the ray Jacobian (up to the sign), and belongs to the plane normal to the central ray.

Figure 2. Initial conditions for the point-source basic solutions: a) Anisotropic medium: Two different positive eigenvalues of matrix $L_{\mathbf{rr}}$, and the zero eigenvalue corresponding to eigenvector $\mathbf{r}$, b) Isotropic medium: Two identical positive eigenvalues and the zero eigenvalue, with locally curved ray trajectory in the proximity of the source. Tangent, normal and bi-normal of the ray trajectory, c) Isotropic medium: Two identical positive eigenvalues and the zero eigenvalue, with locally straight ray trajectory in the proximity of the source. The ray direction vector and the "ghost curvature" vector share the same azimuth: $\psi_S = \psi_{\tilde{S}}$. Angles $\theta_S$ and $\theta_{\tilde{S}} = \theta_S + \pi/2$ are zenith angles of the ray direction and the "ghost curvature", respectively (assuming $\theta_S < \pi/2$). The red arrow is the ray direction, the blue arrow is its projection on the horizontal plane $x_1 x_2$, and the green arrow is the "ghost curvature". These three arrows share the same vertical plane of azimuth $\psi_S$.

Figure 3. Initial conditions for the plane-wave basic solutions: a) Anisotropic medium: Two different positive eigenvalues of matrix $L_{\mathbf{rr}}$, and the zero eigenvalue corresponding to eigenvector $\mathbf{r}$, b) Isotropic medium: Two identical positive eigenvalues and the zero eigenvalue,

with locally curved ray trajectory in the proximity of the source, c) Isotropic medium: Two identical positive eigenvalues and the zero eigenvalue, with locally straight ray trajectory in the proximity of the source. The red arrow is the ray direction, the blue arrow is its projection on the horizontal plane $x_1 x_2$, and the green arrow is the "ghost curvature". These three arrows share the same vertical plane of azimuth $\psi_S$. The shifts at the start point, $\mathbf{u}_{i,S}$, $i = 3, 4$, are normal to the ray and to each other, while their arclength derivatives, $\dot{\mathbf{u}}_{i,S}$, are collinear to the ray direction and proportional to the local curvature of the central ray at the start point. For a locally straight central ray, these derivatives vanish.

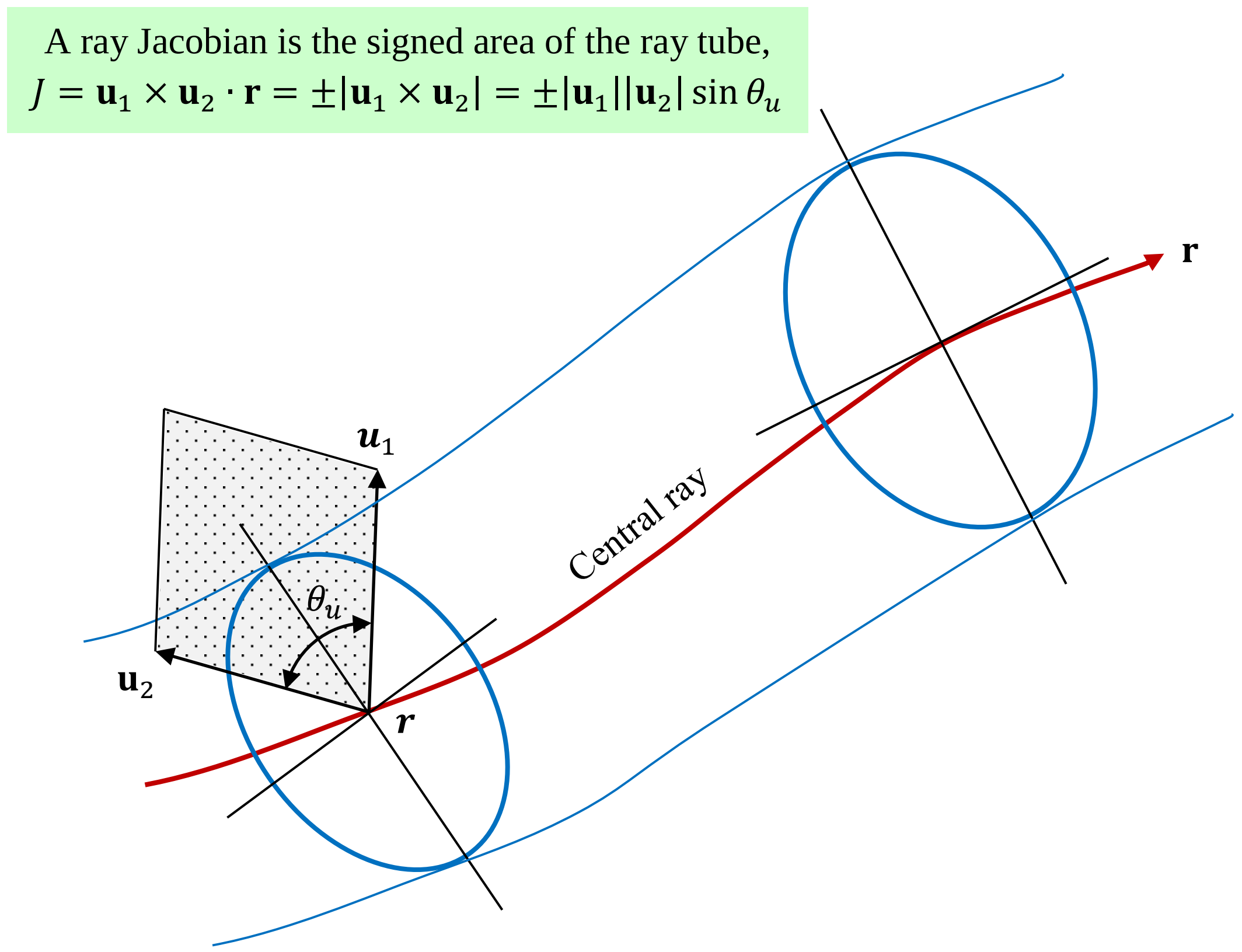

Figure 1. A scheme of a ray tube for point-source paraxial rays: $\mathbf{r}(s)$ – central ray direction, $\mathbf{u}_1(s)$, $\mathbf{u}_2(s)$ – paraxial shift vectors, normal to $\mathbf{r}(s)$, $\theta_u(s)$ – angle between these vectors, $s$ – arclength of the central ray (red line). At the source point, the normal shifts $\mathbf{u}_1$ and $\mathbf{u}_2$ vanish. In the proximity of the source point, these vectors are infinitesimal and normal to each other. At the line caustic, the normal shift vectors may become collinear. The shaded area of the parallelogram is equal to the ray Jacobian (up to the sign), and belongs to the plane normal to the central ray.

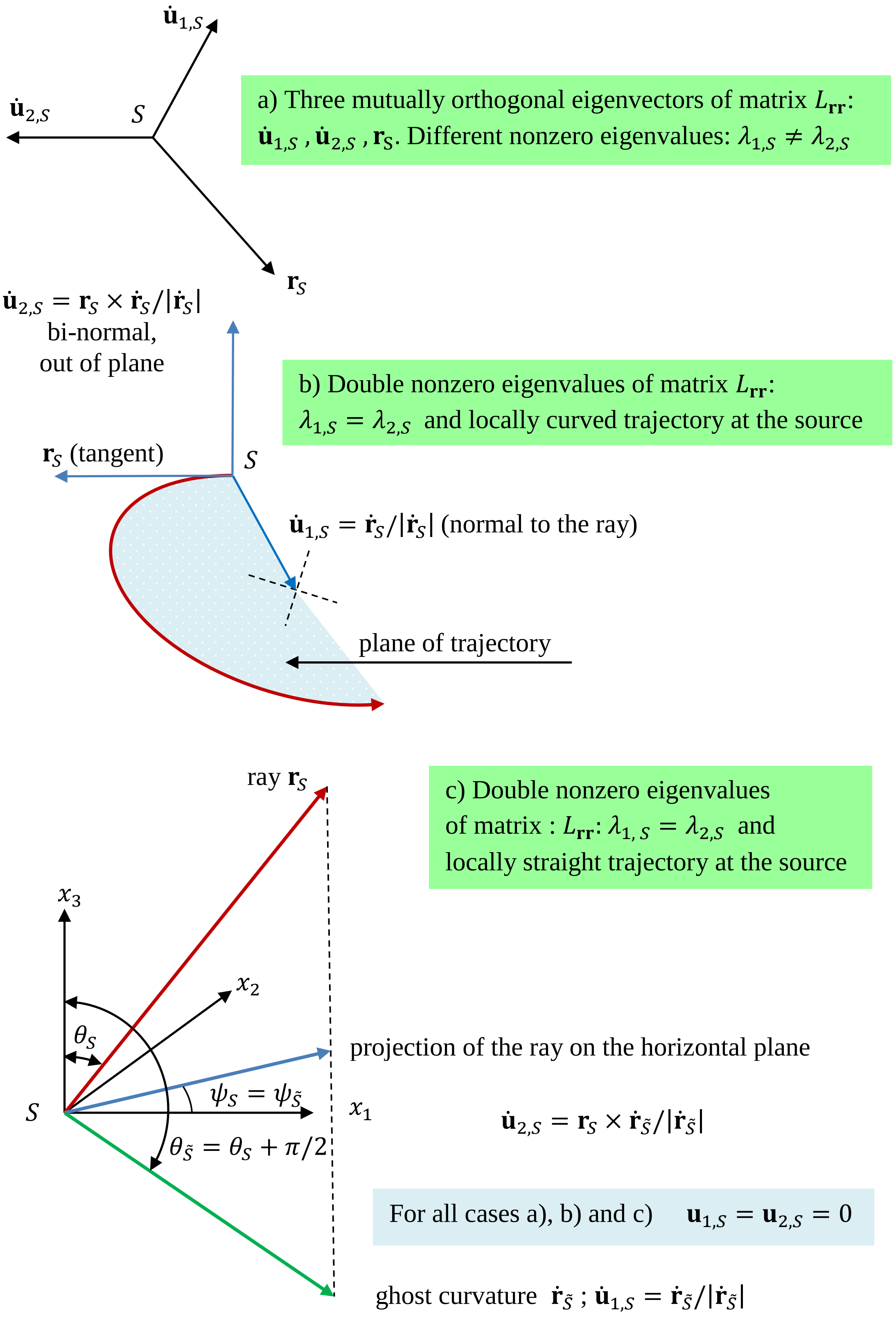
$\dot{\mathbf{u}}_{1,S}$
$\dot{\mathbf{u}}_{2,S}$
$S$
$\mathbf{r}_S$
a) Three mutually orthogonal eigenvectors of matrix $L_{\mathbf{rr}}$:
$\dot{\mathbf{u}}_{1,S}$ , $\dot{\mathbf{u}}_{2,S}$ , $\mathbf{r}_S$. Different nonzero eigenvalues: $\lambda_{1,S} \neq \lambda_{2,S}$
$\dot{\mathbf{u}}_{2,S} = \mathbf{r}_S \times \dot{\mathbf{r}}_S / |\dot{\mathbf{r}}_S|$
bi-normal,
out of plane
b) Double nonzero eigenvalues of matrix $L_{\mathbf{rr}}$:
$\lambda_{1,S} = \lambda_{2,S}$ and locally curved trajectory at the source
$\mathbf{r}_S$ (tangent)
$S$
$\dot{\mathbf{u}}_{1,S} = \dot{\mathbf{r}}_S / |\dot{\mathbf{r}}_S|$ (normal to the ray)
plane of trajectory
ray $\mathbf{r}_S$
c) Double nonzero eigenvalues
of matrix : $L_{\mathbf{rr}}$: $\lambda_{1,S} = \lambda_{2,S}$ and
locally straight trajectory at the source
$x_3$
$x_2$
$\theta_S$
projection of the ray on the horizontal plane
$S$
$\psi_S = \psi_{\tilde{S}}$
$x_1$
$\dot{\mathbf{u}}_{2,S} = \mathbf{r}_S \times \dot{\mathbf{r}}_{\tilde{S}} / |\dot{\mathbf{r}}_{\tilde{S}}|$
$\theta_{\tilde{S}} = \theta_S + \pi/2$
For all cases a), b) and c) $\mathbf{u}_{1,S} = \mathbf{u}_{2,S} = 0$
ghost curvature $\dot{\mathbf{r}}_{\tilde{S}}$ ; $\dot{\mathbf{u}}_{1,S} = \dot{\mathbf{r}}_{\tilde{S}} / |\dot{\mathbf{r}}_{\tilde{S}}|$

Figure 2. Initial conditions for the point-source basic solutions: a) Anisotropic medium: Two different positive eigenvalues of matrix $L_{\mathbf{rr}}$, and the zero eigenvalue corresponding to eigenvector $\mathbf{r}$, b) Isotropic medium: Two identical positive eigenvalues and the zero eigenvalue, with locally curved ray trajectory in the proximity of the source. Tangent, normal and bi-normal of the ray trajectory, c) Isotropic medium: Two identical positive eigenvalues and the zero eigenvalue, with locally straight ray trajectory in the proximity of the source. The ray direction vector and the "ghost curvature" vector share the same azimuth: $\psi_S = \psi_{\tilde{S}}$. Angles $\theta_S$ and $\theta_{\tilde{S}} = \theta_S + \pi / 2$ are zenith angles of the ray direction and the "ghost curvature", respectively (assuming $\theta_S < \pi / 2$). The red arrow is the ray direction, the blue arrow is its projection on the horizontal plane $x_1 x_2$, and the green arrow is the "ghost curvature". These three arrows share the same vertical plane of azimuth $\psi_S$.

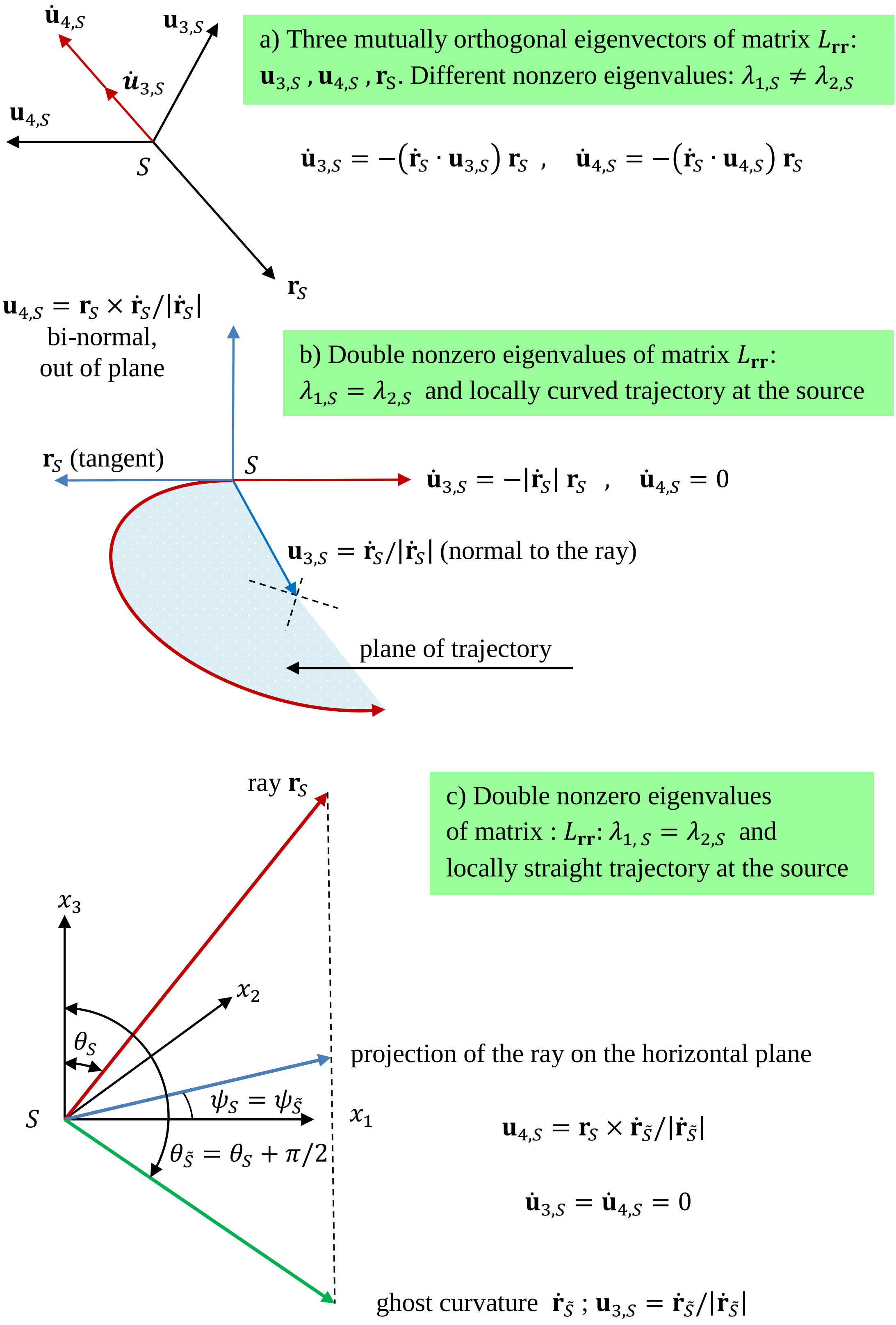
a) Three mutually orthogonal eigenvectors of matrix $L_{\mathbf{rr}}$:
$\mathbf{u}_{3,S}$ , $\mathbf{u}_{4,S}$ , $\mathbf{r}_S$. Different nonzero eigenvalues: $\lambda_{1,S} \neq \lambda_{2,S}$
$\dot{\mathbf{u}}_{4,S}$
$\mathbf{u}_{3,S}$
$\dot{\mathbf{u}}_{3,S}$
$\mathbf{u}_{4,S}$
$S$
$\mathbf{r}_S$
$\dot{\mathbf{u}}_{3,S} = -(\dot{\mathbf{r}}_S \cdot \mathbf{u}_{3,S})\,\mathbf{r}_S$ , $\dot{\mathbf{u}}_{4,S} = -(\dot{\mathbf{r}}_S \cdot \mathbf{u}_{4,S})\,\mathbf{r}_S$
$\mathbf{u}_{4,S} = \mathbf{r}_S \times \dot{\mathbf{r}}_S / |\dot{\mathbf{r}}_S|$
bi-normal,
out of plane
b) Double nonzero eigenvalues of matrix $L_{\mathbf{rr}}$:
$\lambda_{1,S} = \lambda_{2,S}$ and locally curved trajectory at the source
$\mathbf{r}_S$ (tangent)
$S$
$\dot{\mathbf{u}}_{3,S} = -|\dot{\mathbf{r}}_S|\,\mathbf{r}_S$ , $\dot{\mathbf{u}}_{4,S} = 0$
$\mathbf{u}_{3,S} = \dot{\mathbf{r}}_S / |\dot{\mathbf{r}}_S|$ (normal to the ray)
plane of trajectory
ray $\mathbf{r}_S$
c) Double nonzero eigenvalues
of matrix : $L_{\mathbf{rr}}$: $\lambda_{1,S} = \lambda_{2,S}$ and
locally straight trajectory at the source
$x_3$
$x_2$
$\theta_S$
projection of the ray on the horizontal plane
$S$
$\psi_S = \psi_{\tilde{S}}$
$x_1$
$\theta_{\tilde{S}} = \theta_S + \pi/2$
$\mathbf{u}_{4,S} = \mathbf{r}_S \times \dot{\mathbf{r}}_{\tilde{S}} / |\dot{\mathbf{r}}_{\tilde{S}}|$
$\dot{\mathbf{u}}_{3,S} = \dot{\mathbf{u}}_{4,S} = 0$
ghost curvature $\dot{\mathbf{r}}_{\tilde{S}}$ ; $\mathbf{u}_{3,S} = \dot{\mathbf{r}}_{\tilde{S}} / |\dot{\mathbf{r}}_{\tilde{S}}|$

Figure 3. Initial conditions for the plane-wave basic solutions: a) Anisotropic medium: Two different positive eigenvalues of matrix $L_{\mathbf{rr}}$, and the zero eigenvalue corresponding to eigenvector $\mathbf{r}$, b) Isotropic medium: Two identical positive eigenvalues and the zero eigenvalue, with locally curved ray trajectory in the proximity of the source, c) Isotropic medium: Two identical positive eigenvalues and the zero eigenvalue, with locally straight ray trajectory in the proximity of the source. The red arrow is the ray direction, the blue arrow is its projection on the horizontal plane $x_1 x_2$, and the green arrow is the "ghost curvature". These three arrows share the same vertical plane of azimuth $\psi_S$. The shifts at the start point, $\mathbf{u}_{i,S}$, $i = 3,4$, are normal to the ray and to each other, while their arclength derivatives, $\dot{\mathbf{u}}_{i,S}$, are collinear to the ray direction and proportional to the local curvature of the central ray at the start point. For a locally straight central ray, these derivatives vanish.